\documentclass[a4paper]{aa}
\usepackage{amssymb}
\usepackage{amsmath}
\usepackage{graphicx}
\usepackage{txfonts}
\usepackage[bookmarks, colorlinks, breaklinks]{hyperref}
\hypersetup{linkcolor=blue,citecolor=blue,filecolor=black,urlcolor=blue} 
\usepackage{natbib}
\bibpunct{(}{)}{;}{a}{}{,} 
\usepackage[percent]{overpic}

\begin{document} 


\title{Using 3D Voronoi grids in radiative transfer simulations}
\titlerunning{3D Voronoi grids in radiative transfer}

\author{P. Camps \and M. Baes \and W. Saftly}
\authorrunning{P. Camps et al.}

\institute{Sterrenkundig Observatorium, Universiteit Gent, Krijgslaan 281, B-9000 Gent, Belgium\\
                 \email{peter.camps@ugent.be}
             }
\date{\today}

\abstract
   {Probing the structure of complex astrophysical objects requires effective three-dimensional (3D) numerical simulation of the relevant
    radiative transfer (RT) processes. As with any numerical simulation code, the choice of an appropriate discretization is crucial.
    Adaptive grids with cuboidal cells such as octrees have proven very popular, however several
    recently introduced hydrodynamical and RT codes are based on a Voronoi tessellation of the spatial domain.
    Such an unstructured grid poses new challenges in laying down the rays (straight paths) needed in RT codes.}
   {We show that it is straightforward to implement accurate and efficient RT on 3D Voronoi grids.}
   {We present a method for computing straight paths between two arbitrary points through a 3D Voronoi grid in the context of a RT
    code. We implement such a grid in our RT code SKIRT, using the open source library Voro++ to obtain the relevant properties of the
    Voronoi grid cells based solely on the generating points. We compare the results obtained through the Voronoi grid with
    those generated by an octree grid for two synthetic models, and we perform the well-known Pascucci RT benchmark using the Voronoi grid.}
   {The presented algorithm produces correct results for our test models. Shooting photon packages through the
    geometrically much more complex 3D Voronoi grid is only about three times slower than the equivalent process in an octree grid
    with the same number of cells, while in fact the total number of Voronoi grid cells may be lower for an equally good representation
    of the density field.}
   {The benefits of using a Voronoi grid in RT simulation codes will often outweigh the somewhat slower performance.}


\maketitle


\section{Introduction}

\vspace{15 mm}  

The radiation observed from most astrophysical systems has been substantially affected by the gas and/or dust residing 
in or in front of the system under study. Many of these systems have a complicated three-dimensional (3D) geometry,
for example arm structures in spiral galaxies \citep{2000A&A...353..117M, 2012A&A...546A..34F},
filaments and clumps in star-forming regions \citep{2008ApJ...680..428G, 2013MNRAS.433.1619P, 2013arXiv1307.0022F},
and bow-shocks around evolved stars \citep{2012A&A...548A.113D, 2013ApJ...769..122W}.
Properly probing the structure of these systems requires a 3D numerical treatment of the relevant
radiative transfer (RT) processes \citep{1989MNRAS.239..939D, 1992ApJ...393..611W, 2001MNRAS.326..733B},
which may include dust, line, ionizing UV, Ly$\alpha$, neutron and neutrino RT.

Due to the complexities involved in multiple anisotropic scattering, absorption and reemission, the full
radiative transfer problem is highly nonlocal and nonlinear.
It is not feasible to directly integrate the equations in three dimensions except for the simplest problems.
Consequently virtually all modern RT codes use ray tracing or Monte Carlo techniques
\citep{2001ApJ...551..269G, 2001MNRAS.324..381C, 2001A&A...379..336K, 2003A&A...397..201J,
2003CoPhC.150...99W, 2003A&A...407..941S, 2004ApJ...610..801B,
2004MNRAS.350..565H, 2004MNRAS.348.1337W, 2005MNRAS.362..737D, 2005MNRAS.362.1038E,
2006A&A...456....1N, 2006MNRAS.372....2J, 2006A&A...459..797P,
2006A&A...460..397V, 2006ApJ...645..792T, 2006ApJ...645..920S,
2008A&A...490..461B, 2009ApJ...696..853L, 2011ApJS..196...22B, 2011A&A...536A..79R,
2012ApJ...751...27H, 2012ApJ...755..111A, 2012A&A...544A..52L, 2013arXiv1303.4998S}. 
In addition to their intrinsic 3D nature, these techniques allow including, for example, a clumpy 
medium \citep{1996ApJ...463..681W, 2000ApJ...528..799W,
2000MNRAS.311..601B, 2005MNRAS.362..737D, 2012MNRAS.420.2756S,
2012ApJ...746...70S}, polarization \citep{1995ApJ...441..400C,
2007A&A...465..129G} or kinematical information
\citep{2001ApJ...548..150M, 2002MNRAS.335..441B, 2003MNRAS.343.1081B}.
Recent RT applications explicitly using 3D include models of young stellar objects \citep{1998A&A...340..103W},
protostellar to protoplanetary disks \citep{2006ApJ...636..362I, 2006A&A...456....1N},
reflection nebulae \citep{1996ApJ...463..681W},
molecular clouds \citep{2009A&A...502..833P, 2005A&A...434..167S}, 
spiral galaxies \citep{2008A&A...490..461B, 2012ApJ...746...70S, 2012MNRAS.427.2797D}, 
interacting and starburst galaxies \citep{2007ApJ...658..840C, 2011ApJ...743..159H},
and active galactic nuclei \citep{2008A&A...482...67S, 2012MNRAS.420.2756S}. 

For the purpose of numerical computation, the domain under study must be discretized. 
Since memory requirements and computation time rapidly increase with the number of grid cells,
modern 3D RT codes employ an adaptive grid, placing more and smaller cells in areas that require a higher resolution.
Starting from a cuboidal root cell that spans the complete spatial domain, an adaptive mesh refinement (AMR) scheme
recursively subdivides each cell in $k \times l \times m$ cuboidal subcells until the required resolution is reached. In the special
case of an octree, $k=l=m=2$ so that each cell is subdivided in eight subcells (hence the name of the data structure).
AMR-based grids and especially octree grids are well established 
\citep{2001A&A...379..336K, Steinacker2002765, 2003CoPhC.150...99W, 2004MNRAS.350..565H,
2006A&A...456....1N, 2006MNRAS.372....2J, 2008A&A...490..461B, 2009ApJ...696..853L, 2011A&A...536A..79R, 2012A&A...544A..52L,
2012ApJ...751...27H, 2013A&A...554A..10S}
and several methods have been investigated to make them as efficient as possible \citep{2013A&A...554A..10S}.

AMR grids seem to be an obvious choice: it is straightforward to construct an appropriate grid for any density field, 
whether defined by an analytical model or by a collection of smoothed particles;
and it is easy to calculate a straight path through the grid, since the boundaries of the cuboidal cells 
are lined up with the coordinate axes and each cell has a limited number of neighbors \citep{2013A&A...554A..10S}.
This second point is very important in the context of RT because 
ray tracing and Monte Carlo RT codes determine the radiation field in each grid cell by laying down
random rays (i.e. straight paths) through the domain. 
The simulation run time is often dominated by the portion of the code that identifies the 
grid cells crossed by each path and calculates the lengths of the corresponding path segments.

\begin{figure}
  \centering
  \includegraphics[width=0.95\columnwidth]{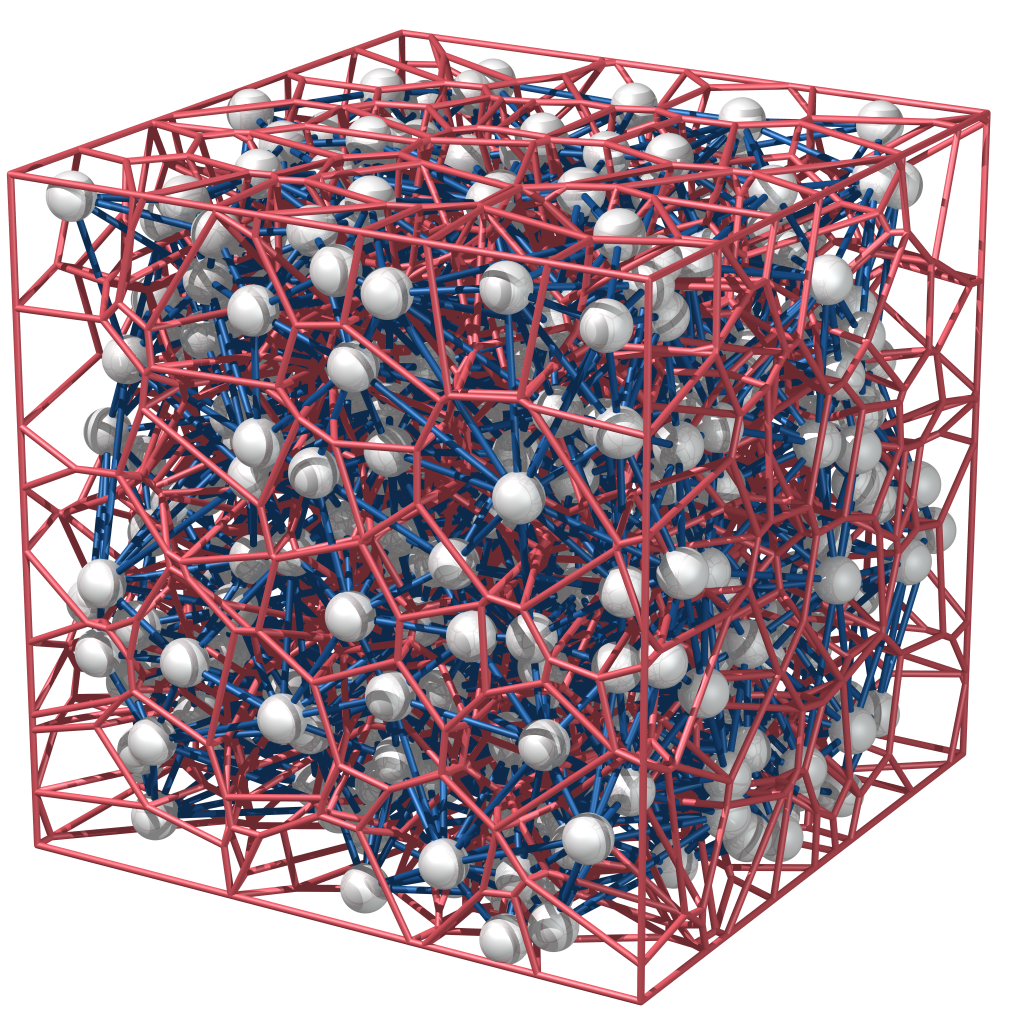}
  \caption{A Voronoi tessellation for 400 random sites (in grey), bounded by a cube. Voronoi cell edges are shown in red, Delaunay edges in blue.}
  \label{fig:tessellation}
\end{figure}

AMR grids also have drawbacks. First of all, for a given density field and required resolution, an AMR grid may not be
the kind of grid with the least number of cells. To illustrate this, consider a density field defined by a set of smoothed particles.
An octree grid constructed such that each cell encloses at most one particle usually has over three times 
more cells than there are particles; i.e. two out of three cells are empty\footnote{To verify this claim, we ran a few 
tests with particles distributed uniformly over the spatial domain, and with particles representing a galaxy generated by a hydrodynamical simulation.}.
In contrast an unstructured grid based on a Voronoi tessellation
of the spatial domain (see Sect.\,\ref{sec:vorodef} and Fig.\,\ref{fig:tessellation}),
using the given particles as generating sites, has exactly the same number of cells as there are particles.
While not an issue in many situations, minimizing the number of cells is sometimes crucial.
For example, consider a panchromatic RT simulation involving small dust grains in non-LTE conditions.
Because each cell stores radiation field data per wavelength bin, memory requirements are substantial.
Moreover the simulation run time is most likely dominated by the calculation  
of the non-LTE heating and re-emission of the dust grains in each cell.
In this case, both memory usage and run time scale roughly linearly with the number of cells.

Furthermore 
RT simulations frequently serve to predict the observable properties of artificial systems resulting from
\mbox{(magneto-)hydrodynamical}  (MHD) simulations \citep{2003A&A...397..201J,
2004ApJ...610..801B, 2005A&A...439..153S, 2010MNRAS.403...17J,
2010MNRAS.403.1143A, 2011ApJ...743..159H, 2011A&A...536A..79R,
2012A&A...544A..52L, 2012A&A...544A.141J}.
MHD simulation codes historically employ one of two schemes: 
a Lagrangian formulation based on moving particles (Smoothed Particle Hydrodynamics or SPH),
for example Gadget \citep{2005MNRAS.364.1105S, 2009MNRAS.398.1678D, 2012MNRAS.424.2222P}
and SEREN \citep{2011A&A...529A..27H};
and a Eulerian approach based on a non-moving spatial grid, often an AMR grid,
for example RAMSES \citep{2006A&A...457..371F}, Enzo \citep{2010ApJS..186..308C, 2013arXiv1307.2265T},
and AMR-VAC \citep{Keppens2012718}.

Recent codes including TESS \citep{2011ApJS..197...15D} and 
AREPO \citep{2010MNRAS.401..791S, 2011arXiv1109.2218S} introduce
a new scheme that employs a moving mesh based on a Voronoi tessellation
of the spatial domain (see Sect.\,\ref{sec:vorodef} and Fig.\,\ref{fig:tessellation}).
This new scheme is claimed to combine the best features of SPH and the traditional Eulerian approach,
and it is becoming increasingly popular.
It has already been applied to various problems including the formation of stars, galaxies and cosmological structure
\citep{2011ApJ...737...75G, 2012MNRAS.423.2558B, 2012MNRAS.424.2999S, 2012MNRAS.425.2027K,
2012MNRAS.425.3024V, 2012MNRAS.427.2224T, 2013MNRAS.429.3353N, 2013arXiv1305.5360M}.
While the output from a moving mesh code can be re-gridded to an AMR grid to perform RT,
the resampling process unavoidably introduces inaccuracies and represents additional overhead;
one would rather perform both aspects of the simulation (MHD and RT) on the same grid.


\begin{figure}
  \centering
  \includegraphics[width=0.95\columnwidth]{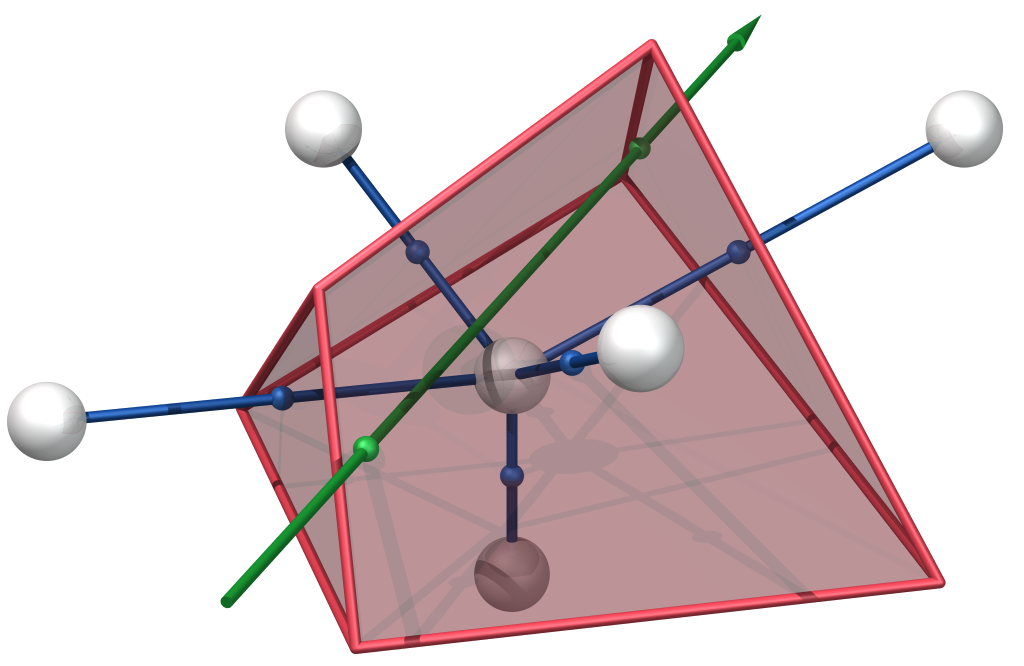}
  \caption{A single Voronoi cell (in red) with its neighboring sites (in grey) and corresponding Delaunay edges (in blue). 
  A straight path through the cell is shown (in green) with its intersection points with the cell boundary at entry and exit.}
  \label{fig:cell}
\end{figure}

\begin{figure*}
  \centering
  \setlength\fboxsep{1 pt}
  \setlength\fboxrule{0 pt}
  \fbox{\begin{overpic}[width=0.163\textwidth]{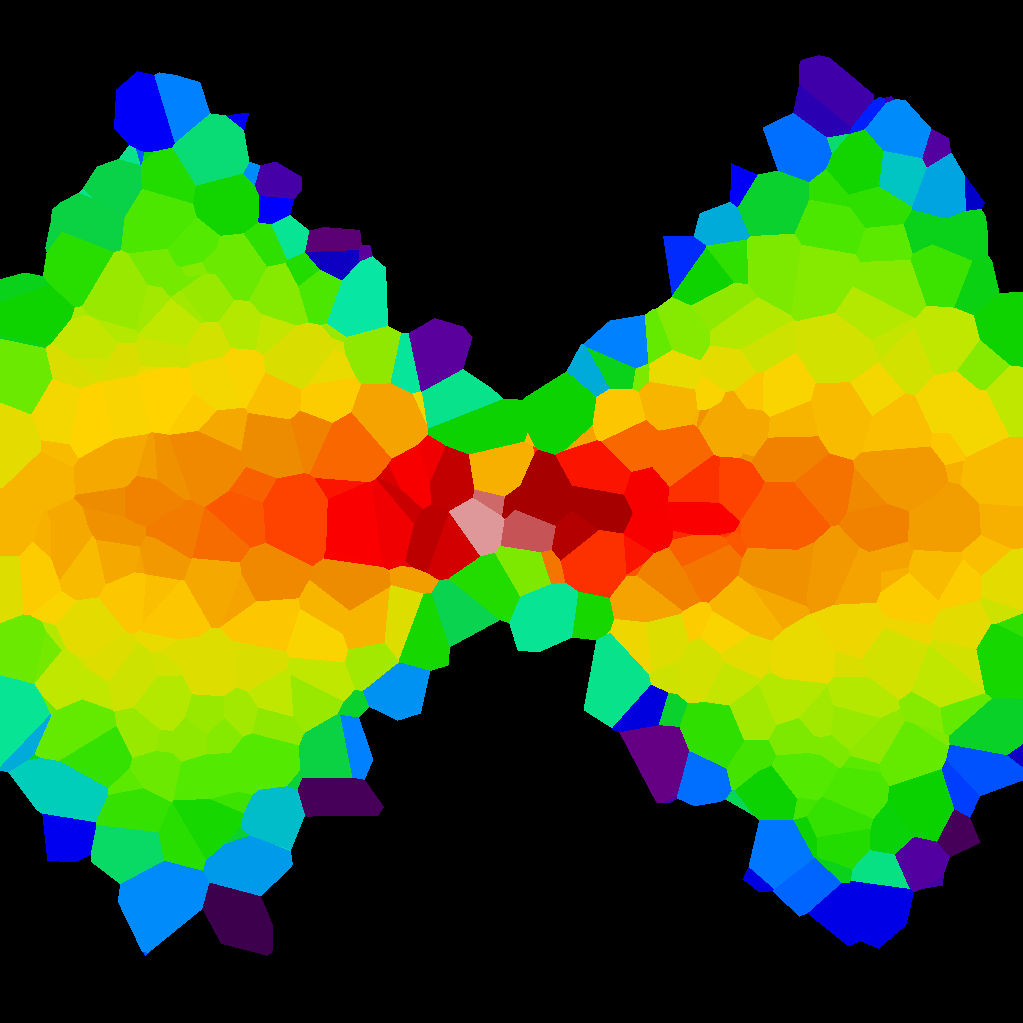} \put(30,89){\color{white} $10^4$ \small cells} \end{overpic}}%
  \fbox{\begin{overpic}[width=0.163\textwidth]{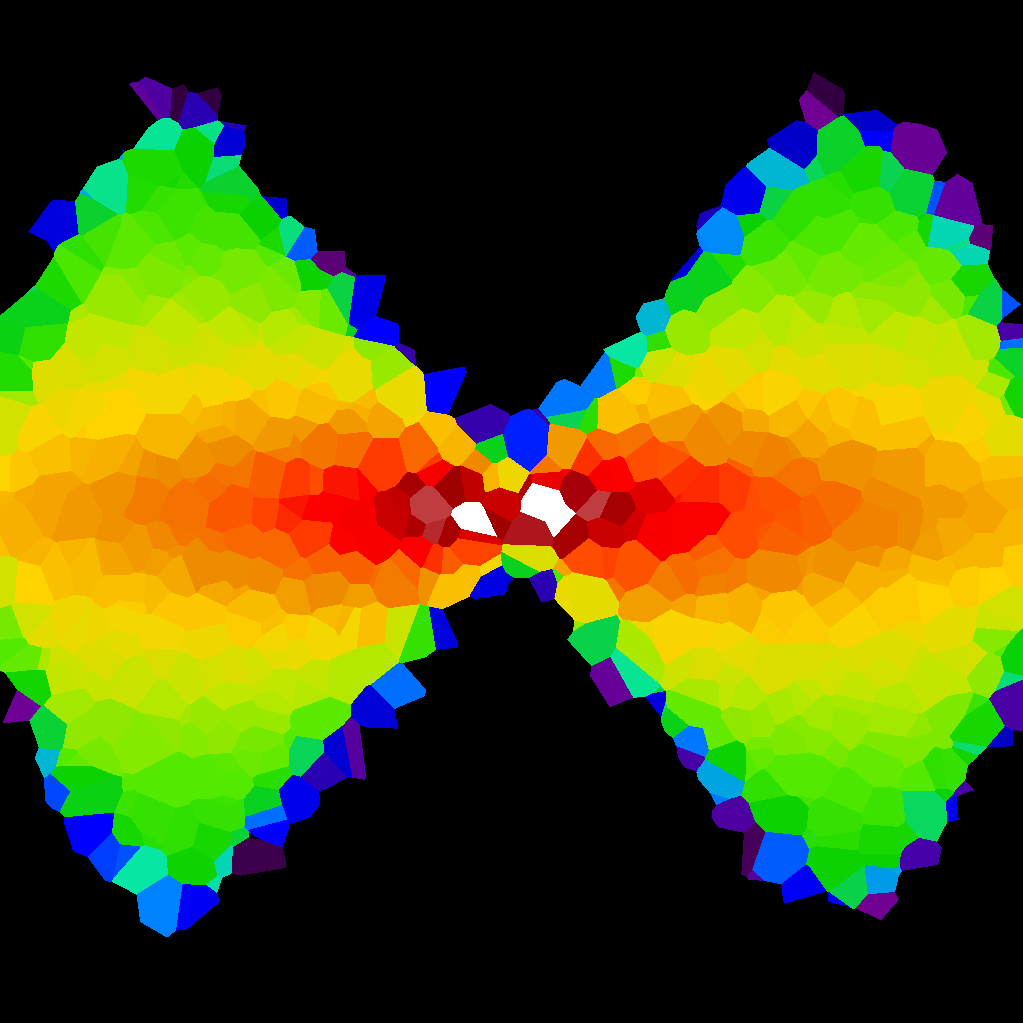} \put(29,89){\color{white} $10^{4.5}$ \small cells} \end{overpic}}%
  \fbox{\begin{overpic}[width=0.163\textwidth]{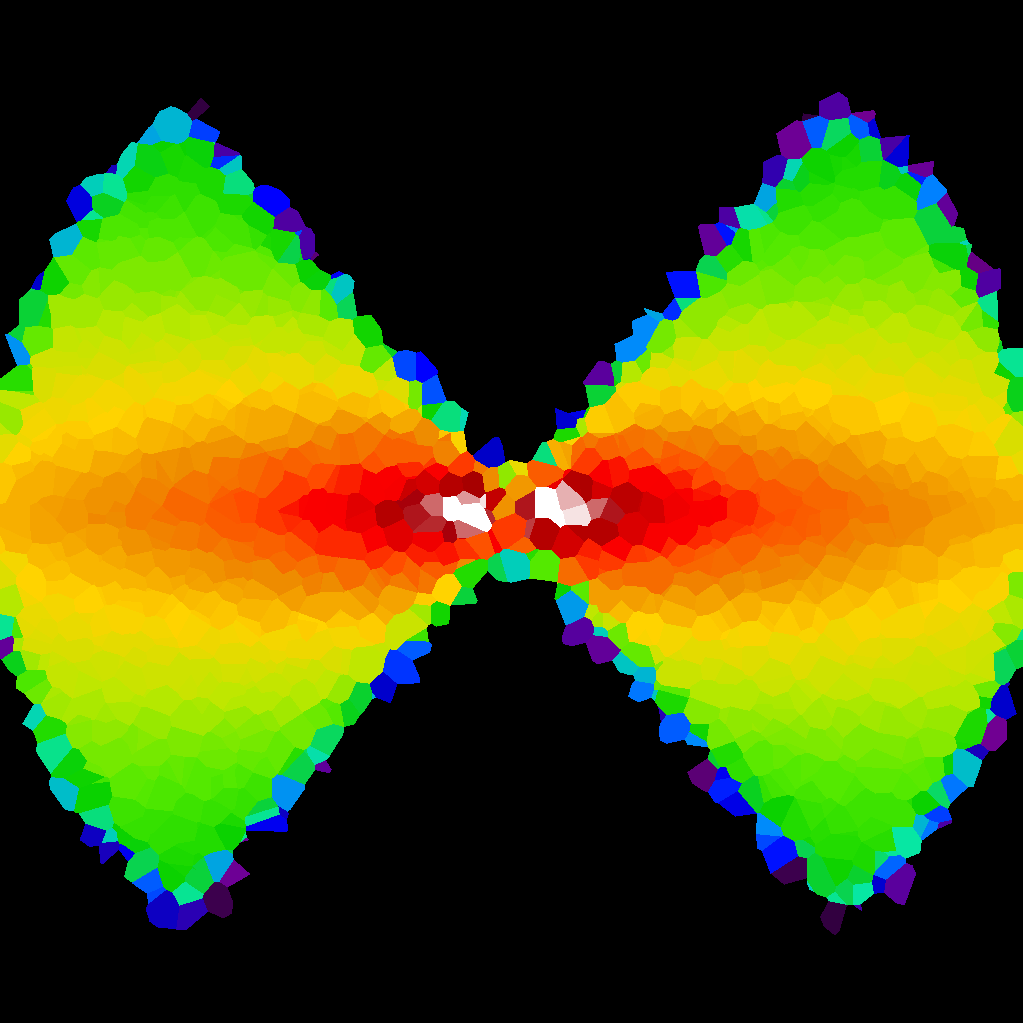} \put(30,89){\color{white} $10^5$ \small cells} \end{overpic}}%
  \fbox{\begin{overpic}[width=0.163\textwidth]{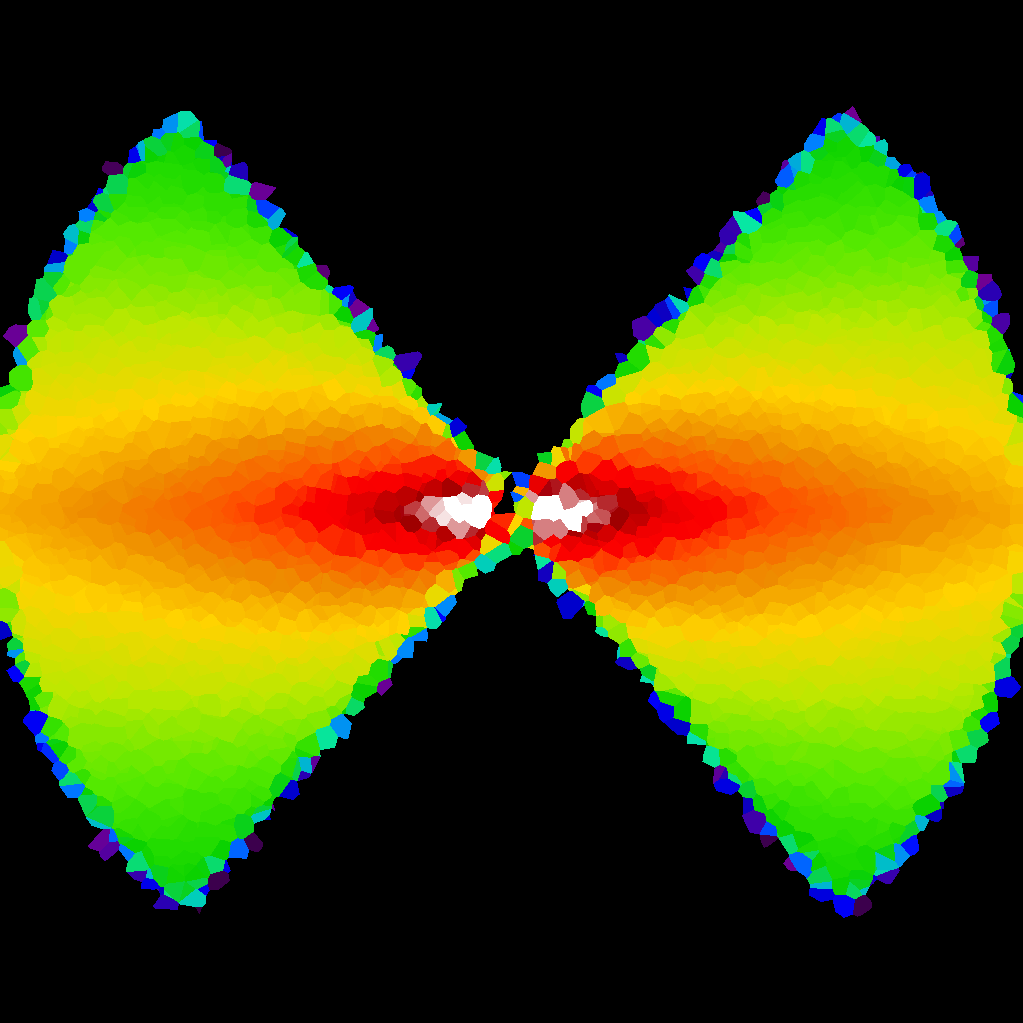} \put(29,89){\color{white} $10^{5.5}$ \small cells} \end{overpic}}%
  \fbox{\begin{overpic}[width=0.163\textwidth]{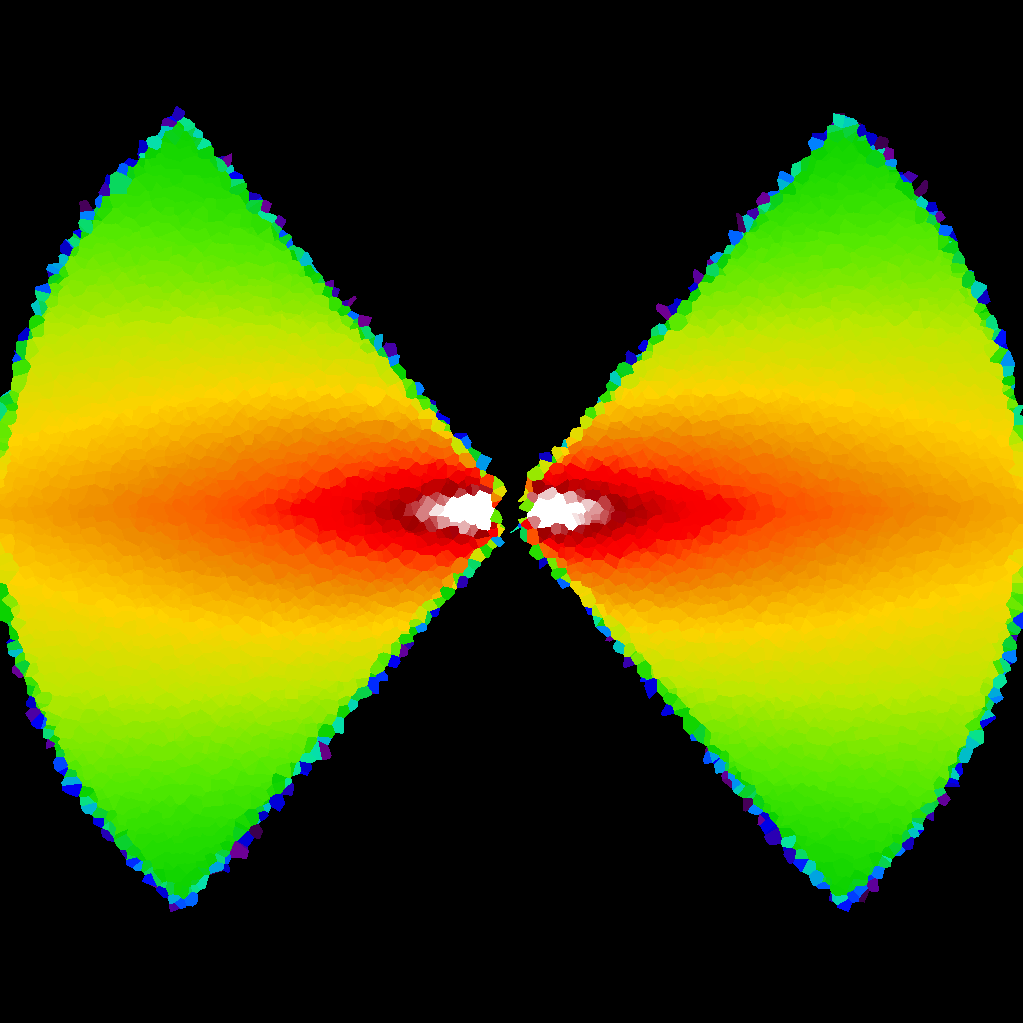} \put(30,89){\color{white} $10^6$ \small cells} \end{overpic}}%
  \fbox{\begin{overpic}[width=0.163\textwidth]{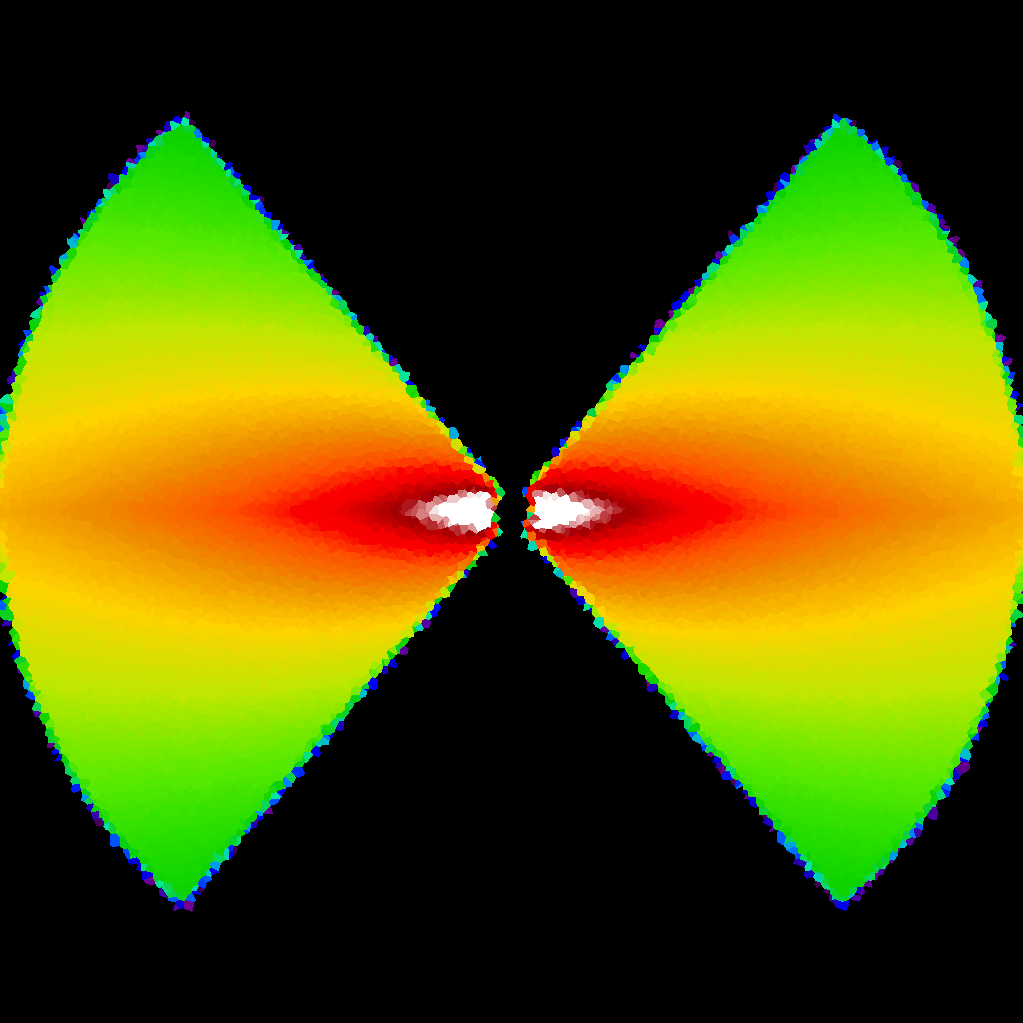} \put(29,89){\color{white} $10^{6.5}$ \small cells} \end{overpic}}%
  \caption{A cut through the dust density distribution of the torus model, discretized on Voronoi grids with a resolution
   varying from $10^4$ cells (\emph{left}) to $10^{6.5}$ cells (\emph{right}).
   All grids were constructed from a set of uniformly distributed sites.}
  \label{fig:resolution}
\end{figure*}


These considerations lead to the question whether it is possible to perform accurate and efficient RT on unstructured Voronoi grids.

One approach is to approximate a straight path through the grid by a sequence of non-collinear segments
connecting neighboring sites. For example in the SimpleX code \citep{2010A&A...515A..79P} and in the LIME code \citep{2010A&A...523A..25B}
radiation travels along the edges of the Delaunay triangulation corresponding to the Voronoi grid (see Sect.\,\ref{sec:vorodef};
the Delaunay edges are shown in blue in Fig.\,\ref{fig:tessellation}).
While it facilitates calculating the paths,
this approximation requires additional mechanisms to compensate for errors in path length (see Fig.\,5 in \citealt{2010A&A...515A..79P}) 
and direction (see Fig.\,4 in \citealt{2010A&A...523A..25B}). The distance covered by the path inside a particular grid cell becomes a fuzzy concept,
while this is an important quantity in many RT codes, e.g.\ for tracking the amount of energy absorbed in the cell.
And finally the spread on direction makes it hard to produce high-resolution images of the simulated object.

In Sect.\,\ref{sec:method} we present instead
an efficient method of calculating a straight path between two arbitrary points through a 3D Voronoi grid, applicable in any RT code based
on ray tracing or Monte Carlo techniques.
The path segments inside each grid cell are calculated to high precision
using a straightforward algorithm that relies on the mathematical properties of Voronoi tessellations.
In Sect.\,\ref{sec:testing} we introduce an implementation of the
method in our dust RT code SKIRT \citep{2011ApJS..196...22B}. We demonstrate the method's reliability, accuracy and efficiency   
by comparing results obtained through the Voronoi grid with those generated by existing well-tested grids.
In Sect.\,\ref{sec:conclusions} we summarize our conclusions.

\section{Method}
\label{sec:method}

\subsection{Voronoi tesselations of 3D space}
\label{sec:vorodef}

Given a set of points $\{\vec{p}_1,\vec{p}_2,\ldots\vec{p}_n\}$ in 3D space, called \emph{sites}, the corresponding Voronoi tessellation
\citep{crll.1850.40.209,crll.1908.134.198} is a set of cells $\{C_i\}$ where each cell $C_i$ consists of all the points $\vec{p}$ at least
as close to $\vec{p}_i$ as to any other site.
The corresponding Delaunay triangulation \citep{1934CSMN.7.793} is a graph created by placing a straight edge between any two sites that
share a cell boundary in the Voronoi tessellation. Thus every site is connected to its nearest neighbors.

An example Voronoi tessellation is shown in Fig.\,\ref{fig:tessellation}, and a single Voronoi cell is illustrated in Fig.\,\ref{fig:cell}.
A Voronoi cell is delimited by a convex polyhedron. A Delaunay edge, i.e. a line segment that connects two sites sharing a polygonal face,
is perpendicularly bisected by the plane containing the face, although the bisection point may lie outside the face.
For a set of sites chosen randomly from a uniform distribution, the number of nearest neighbors (or equivalently the number of cells sharing a face with
any given cell) has an expectation value of  $15.54$ \citep{1994A&A...283..361V}.

To obtain an optimal grid in the context of a RT simulation, the Voronoi sites should be more densely packed (generating smaller cells)
in regions where a higher resolution is desired. For example one could select the positions randomly from a probability distribution that follows the density 
of the RT medium, and perhaps place extra sites near sharp edges or large gradients. If the density field is defined by a set of smoothed particles, the
particle locations form natural Voronoi sites. And of course if the density field is already defined on a Voronoi mesh, the original site locations can be used.

\subsection{A straight path through a Voronoi grid}
\label{sec:straight}

Consider a cuboidal spatial domain $\mathcal{D}$ defined by its corner points $(\vec{D}_\mathrm{min},\vec{D}_\mathrm{max})$,
and a set of sites $\{\,\vec{p}_m\in\mathcal{D},\,m=1\ldots M\,\}$. All sites are inside the domain, and the corresponding Voronoi cells are
clipped by the domain walls, as illustrated in Fig.\,\ref{fig:tessellation}. 
Given a ray describing the path of a photon package, defined by a starting point $\vec{r_0}\in\mathcal{D}$ and a direction $\vec{k}$,
our aim is to calculate the ray's consecutive
intersection points with the Voronoi cell walls -- or equivalently, the distance travelled in each cell -- until the ray leaves the domain.
This is illustrated in Fig.\,\ref{fig:cell} for a single cell.
The presented method can easily be adjusted for other domain geometries, or for rays that originate outside the domain.

During a setup phase, before any straight paths are calculated, the following data is prepared:
\begin{enumerate}
\item The domain boundaries $(\vec{D}_\mathrm{min},\vec{D}_\mathrm{max})$.
\item The positions of the sites $\{\,\vec{p}_m,\,m=1\ldots M\,\}$.
\item For each site $\vec{p}_n$, the indices $\{\,m_{n,i},\,i=1\ldots I_n\,\}$ of all sites neighboring that site, 
       or equivalently of all cells neighboring the cell corresponding to that site.
       Domain walls are represented by special (e.g.\ negative) index values.
\end{enumerate}

Data items (1) and (2) are externally specified as part of the problem definition.
The neighbor lists (3) can be easily derived from a Voronoi tessellation or Delaunay triangulation for the specified set of sites,
since nearest neighbor information is the defining characteristic of these concepts.
\emph{No information is needed on the vertices, edges or faces of the polyhedra delimiting the Voronoi cells.}

To begin calculating a straight path, the current point $\vec{r}$ is set to the starting point, 
and the current cell index $m_r$ is set to the index of the cell containing the starting point.
By definition of a Voronoi tessellation, finding the cell containing a given point $\vec{r}$ is equivalent to locating the site $\vec{p}_i$ nearest to $\vec{r}$.
This is a straightforward operation that can easily be optimized as described later in Sect.\,\ref{sec:findcell}.
For the time being assume that there is a function $\mathcal{C}(\vec{r})$ that returns the index $m$ of the cell containing a given point.

Once initialized, the method loops over the algorithm that computes the exit point
from the current cell, i.e. the intersection of the ray formed by the current point $\vec{r}$ and
the direction $\vec{k}$ with the current cell's boundary. The
algorithm also produces the index of the neighboring cell without extra cost. If an exit point
is found, the loop adds a path segment to the output, updates the current point and the
current cell index, and continues to the next iteration. If the exit is towards a domain wall, the loop is terminated.
Due to computational inaccuracies it may occur that no exit point is found. In that case, no path
segment is added to the output, the current point is advanced in the direction $\vec{k}$ over an infinitesimal distance
$\epsilon\ll ||\vec{D}_\mathrm{max}-\vec{D}_\mathrm{min}||$
and the new current cell index is determined by calling the function $\mathcal{C}(\vec{r})$.

The algorithm computing the exit point from the current cell requires the following input data:
the current point $\vec{r}$;
the ray's direction $\vec{k}$ as a unit vector;
the index $m_r$ of the current cell;
the indices $\{\,m_i,\,i=1\ldots I\,\}$ of all cells neighboring the current cell, with domain walls represented by special (e.g.\ negative) values;
the positions of the sites $\{\,\vec{p}_m,\,m=1\ldots M\,\}$;
and the domain boundaries $(\vec{D}_\mathrm{min},\vec{D}_\mathrm{max})$.

The line containing the ray under consideration can be written as $\mathcal{L}(s)=\vec{r}+s\,\vec{k}$ with $s\in\mathbb{R}$. 
The exit point can similarly be written as $\vec{q}=\vec{r}+s_q\,\vec{k}$ with $s_q>0$, and the distance covered within
the cell is given by $s_q$. The index of the cell next to the exit point is denoted $m_q$
and is easily determined as follows.
\begin{enumerate}
\item Calculate the set of values $\{s_i\}$ for the intersection points between the line
           $\mathcal{L}(s)$ and the planes defined by the neighbors $m_i$
           (see below for details on this calculation).
\item Select the smallest nonnegative value $s_q=\min\{s_i|s_i>0\}$ in the set to determine
           the exit point and the corresponding neighbor $m_q$.
\item If there is no nonnegative value in the set, no exit point has been found.
\end{enumerate}

To calculate $s_i$ in step (1) for a regular neighbor $m_i>0$, intersect the
line $\mathcal{L}(s)=\vec{r}+s\,\vec{k}$ with the plane bisecting the sites
$\vec{p}(m_i)$ and $\vec{p}(m_r)$. An unnormalized vector perpendicular to
this plane is given by \begin{equation}\vec{n}=\vec{p}(m_i)-\vec{p}(m_r)\end{equation}
and a point on the plane is given by \begin{equation}\vec{p}=\frac{\vec{p}(m_i)+\vec{p}(m_r)}{2}.\end{equation}
The equation of the plane can then be written as \begin{equation}\vec{n}\cdot(\vec{x}-\vec{p})=0.\label{eq:plane}\end{equation}
Substituting $\vec{x}=\vec{r}+s_i\,\vec{k}$ and solving for $s_i$ provides
\begin{equation}s_i=\frac{\vec{n}\cdot(\vec{p}-\vec{r})}{\vec{n}\cdot\vec{k}}.\label{eq:si}\end{equation}
If $\vec{n}\cdot\vec{k}=0$ the line and the plane are parallel so that there is no intersection, and
the above equation produces $s_i=\pm\infty$. When using standard IEEE 754 floating point arithmetic there is
no reason to test for this special case, since the infinite value will never be selected as the exit point in step (2).

To calculate $s_i$ in step (1) for a domain wall $m_i<0$, substitute the appropriate normal and
position vectors for the wall plane in Eq.\,\ref{eq:si}. For example, for the left wall
one has $\vec{n}=(-1,0,0)$ and $\vec{p}=(D_\mathrm{min,x},0,0)$
so that \begin{equation}s_i=\frac{D_\mathrm{min,x}-r_\mathrm{x}}{k_\mathrm{x}}.\end{equation}

In an actual implementation of this algorithm there is no need to accumulate the complete set of $\{s_i\}$ values; one can simply keep track
of the smallest nonnegative value. As a further optimization, part of the intersection calculation can be avoided for about half of the planes by noting
that the sign of $\vec{n}\cdot\vec{k}$ determines the sign of $s_i$ in Eq.\,\ref{eq:si}. 
Indeed, the site $\vec{p}(m_r)$ and the current point $\vec{r}$
are on the same side of the plane defined by Eq.\,\ref{eq:plane} (unless $\vec{r}$ lies \emph{in} the plane), so that the numerator of Eq.\,\ref{eq:si}
is always positive (or zero if $\vec{r}$ lies in the plane).

\subsection{Finding the cell containing a given point}
\label{sec:findcell}

We now return to the implementation of the function $\mathcal{C}(\vec{r})$ that identifies the Voronoi cell containing a given query point.
By definition of a Voronoi tessellation, this operation is equivalent to finding the site closest to the query point.
There are many sophisticated ways to accelerate this nearest neighbor search, for example by building a
kd-tree \citep{Friedman:1977:AFB:355744.355745} or an R-tree \citep{Guttman:1984:RDI:971697.602266} data structure.
We chose to use a simple mechanism, since this function is usually invoked only once per path and thus its performance is not overly critical.

Assume the domain $\mathcal{D}$ is partitioned in a set of cuboidal \emph{blocks} $\{\,\mathcal{B}_k,\,k=1\ldots K\,\}$ according to a regular linear grid.
During the setup phase described in the beginning of Sect. \ref{sec:straight}, an additional data structure is constructed containing,
for each block $\mathcal{B}_k$ in the partition of the domain $\mathcal{D}$,
the indices $\{\,m_{k,j},\,j=1\ldots J_k\,\}$ of all Voronoi cells that possibly overlap that block.
Determining these lists in principle requires an intersection test between each block and each cell.
In practice it suffices to consider the cell's bounding box, which can be easily intersected with the blocks.

One might be tempted to derive a Voronoi cell's bounding box from the positions of the neighboring sites;
however the convex hull of a cell's neighboring sites does not necessarily fully enclose the cell.
Because a Voronoi cell is convex, its bounding box can be easily calculated from the list of its vertices.
This requires fully constructing the cell, however this is needed anyway to calculate the cell volume 
for use in other areas of the RT simulation (e.g.\ determining the specific energy absorbed per unit mass by the medium in the cell).
In any case the cell geometry is needed solely during setup and does not have to be retained thereafter.

The function $\mathcal{C}(\vec{r})$ receives the following input data:
the query point $\vec{r}$;
for each block $\mathcal{B}_k$, the indices $\{\,m_{k,j},\,j=1\ldots J_k\,\}$ of all Voronoi cells that possibly overlap that block;
the positions of the sites $\{\,\vec{p}_m,\,m=1\ldots M\,\}$;
and the domain boundaries $(\vec{D}_\mathrm{min},\vec{D}_\mathrm{max})$.

For each query, the function $\mathcal{C}(\vec{r})$ performs these steps:
\begin{enumerate}
\item Verify that the query point is inside the domain; if not return a special index, e.g.\ a negative value.
\item Locate the block containing the query point; this is trivial since blocks are on a regular linear grid.
\item Retrieve the list of cells possibly overlapping that block, and thus possibly containing the query point.
\item Calculate the squared distance from the query point to the sites for each of these cells. By definition of a Voronoi tesselation,
        the closest site determines the Voronoi cell containing the query point.
\end{enumerate}


\begin{figure*}
  \centering 
  \setlength\fboxsep{1 pt}
  \setlength\fboxrule{0 pt}
  \fbox{\begin{overpic}[width=0.32\textwidth]{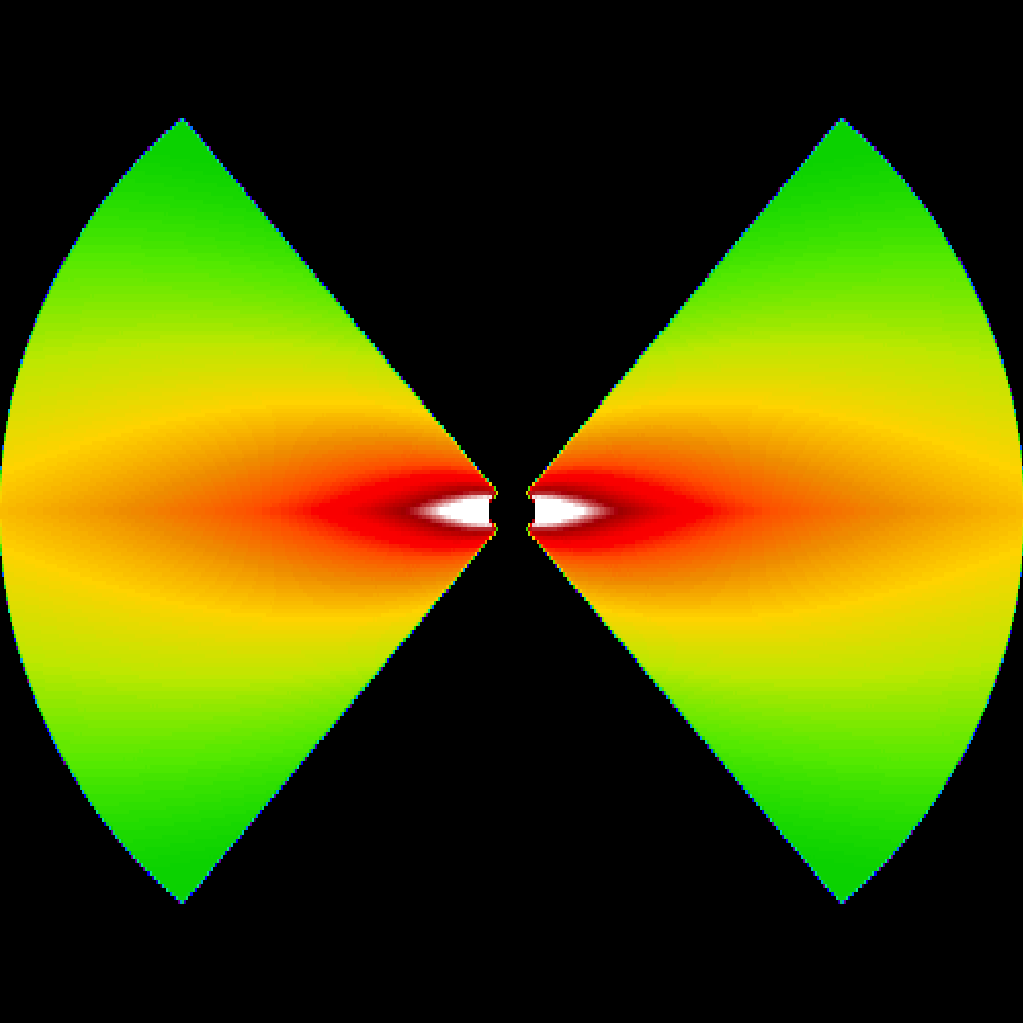} \put(28,93){\color{white} $\rho_\mathrm{dust}$ \small -- cylindrical grid} \end{overpic}}%
  \fbox{\begin{overpic}[width=0.32\textwidth]{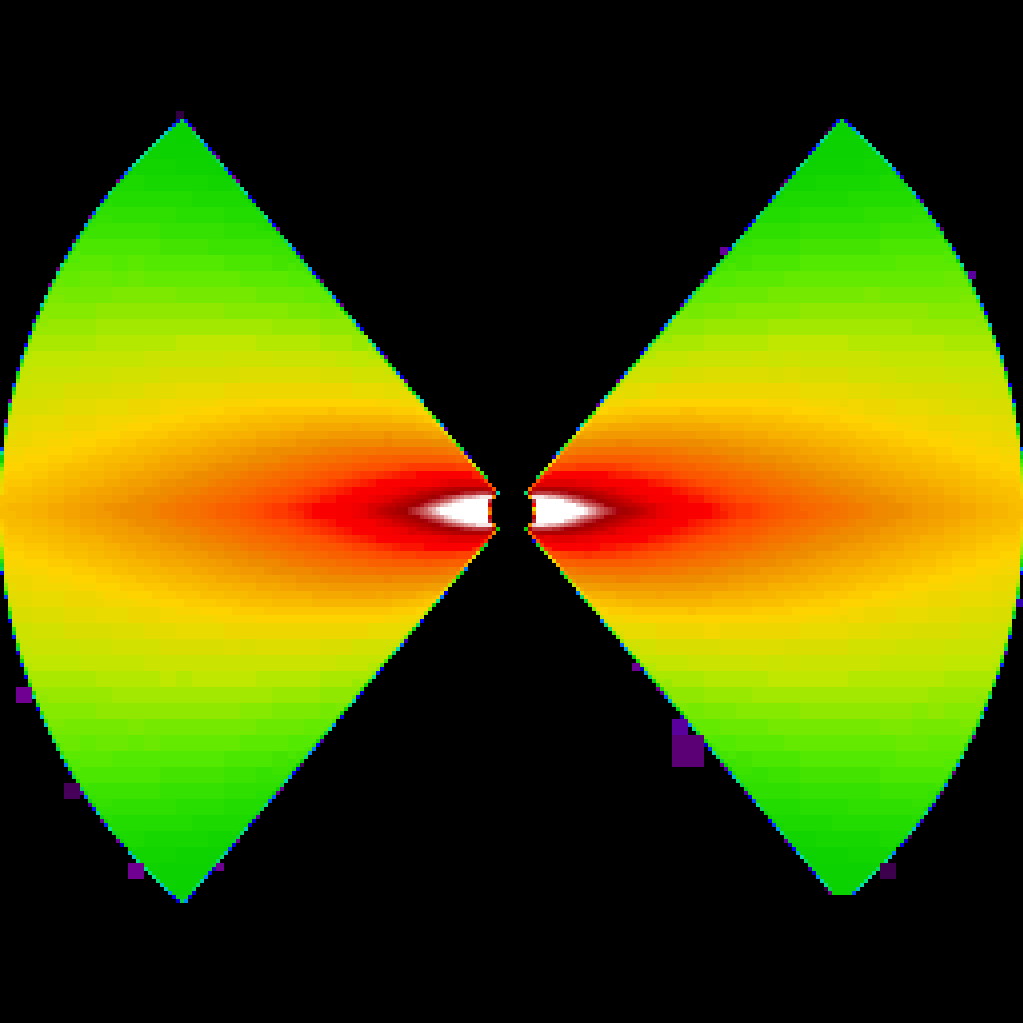} \put(32,93){\color{white} $\rho_\mathrm{dust}$ \small -- octree grid} \end{overpic}}%
  \fbox{\begin{overpic}[width=0.32\textwidth]{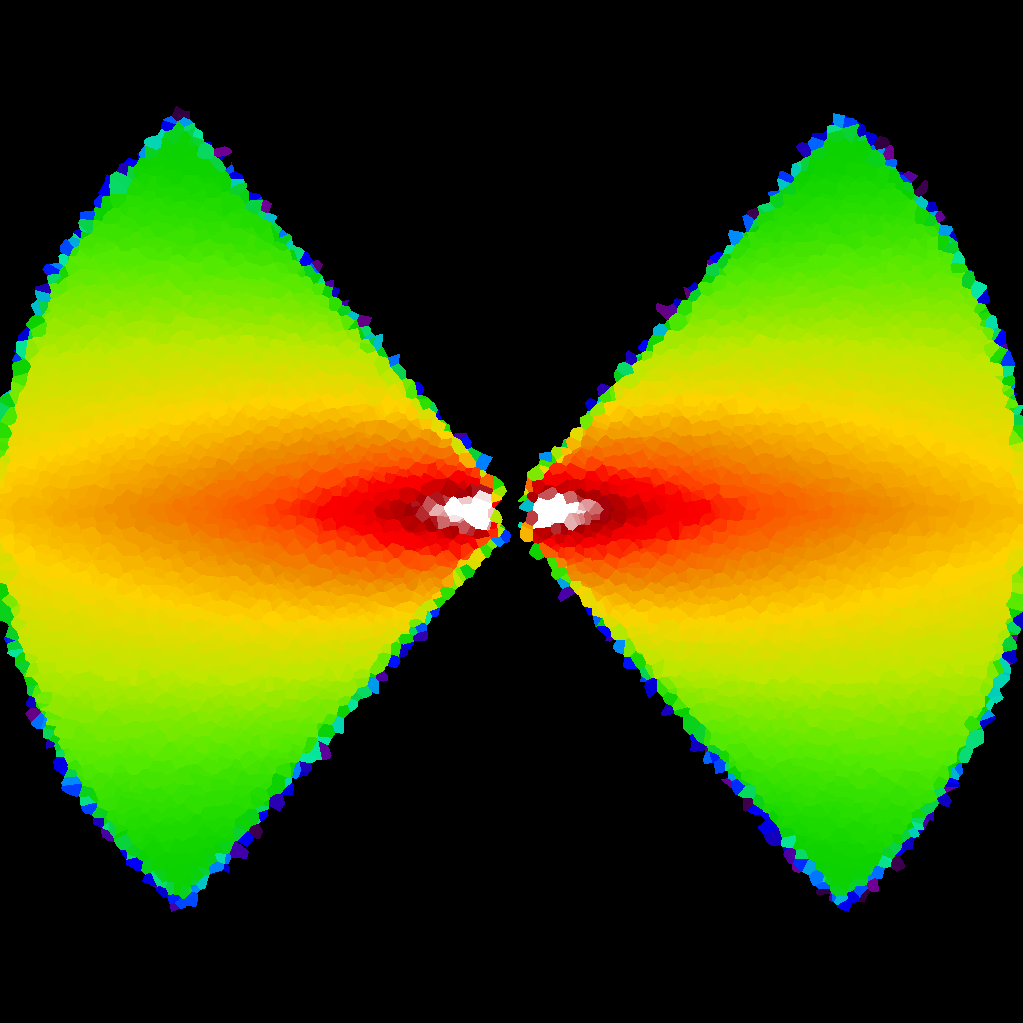} \put(30,93){\color{white} $\rho_\mathrm{dust}$ \small -- Voronoi grid} \end{overpic}}%
  \fbox{\begin{overpic}[height=0.32\textwidth]{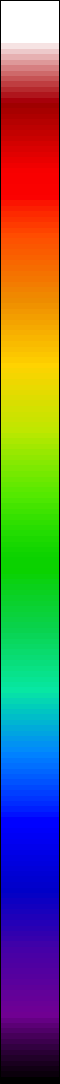} \end{overpic}}\\
  \fbox{\begin{overpic}[width=0.32\textwidth]{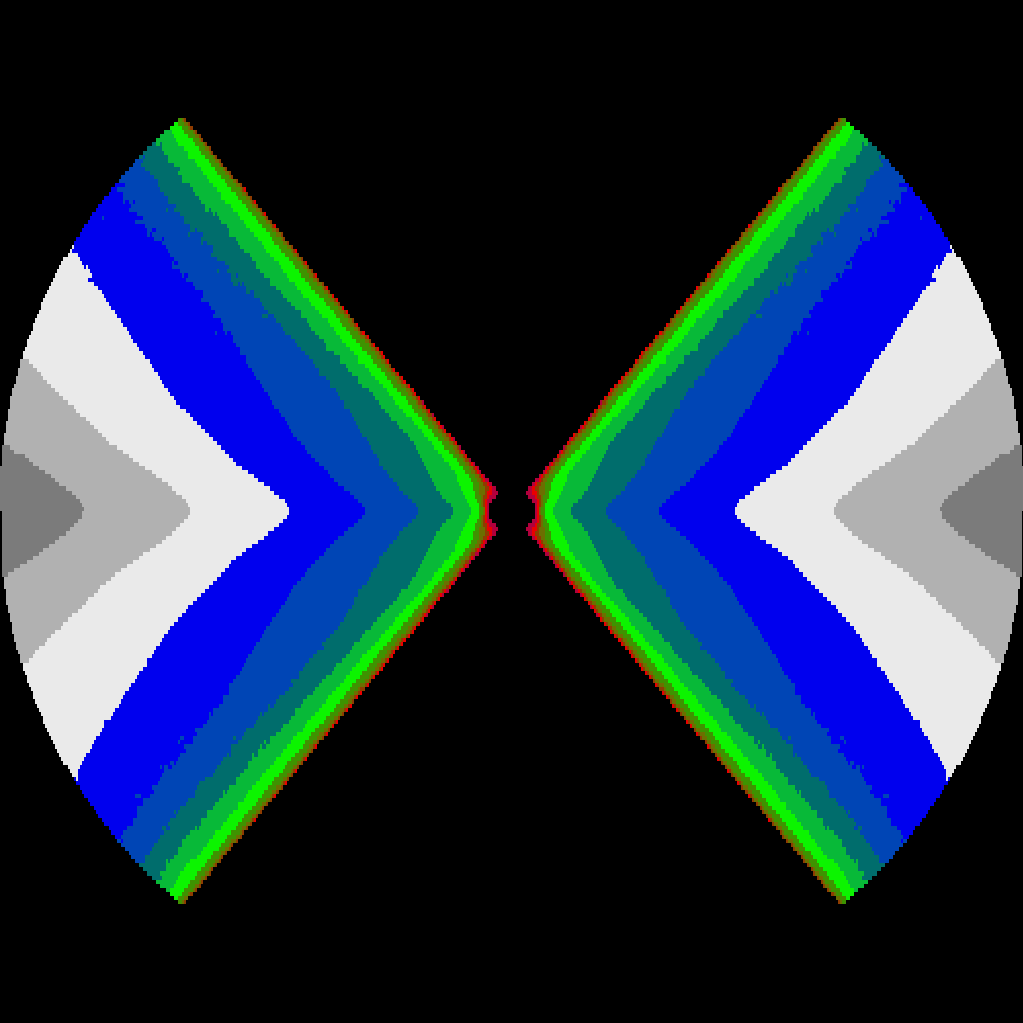} \put(28,93){\color{white} $T_\mathrm{dust}$ \small -- cylindrical grid} \end{overpic}}%
  \fbox{\begin{overpic}[width=0.32\textwidth]{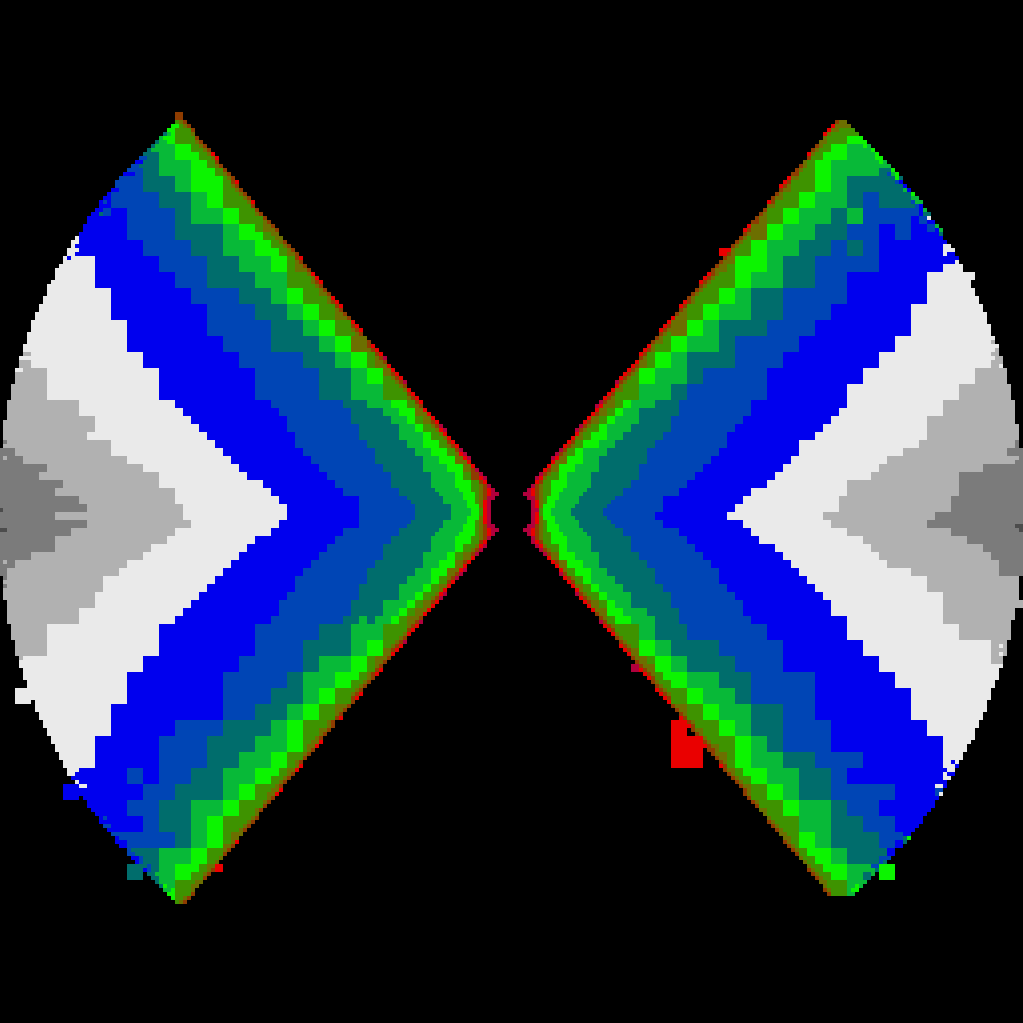} \put(32,93){\color{white} $T_\mathrm{dust}$ \small -- octree grid} \end{overpic}}%
  \fbox{\begin{overpic}[width=0.32\textwidth]{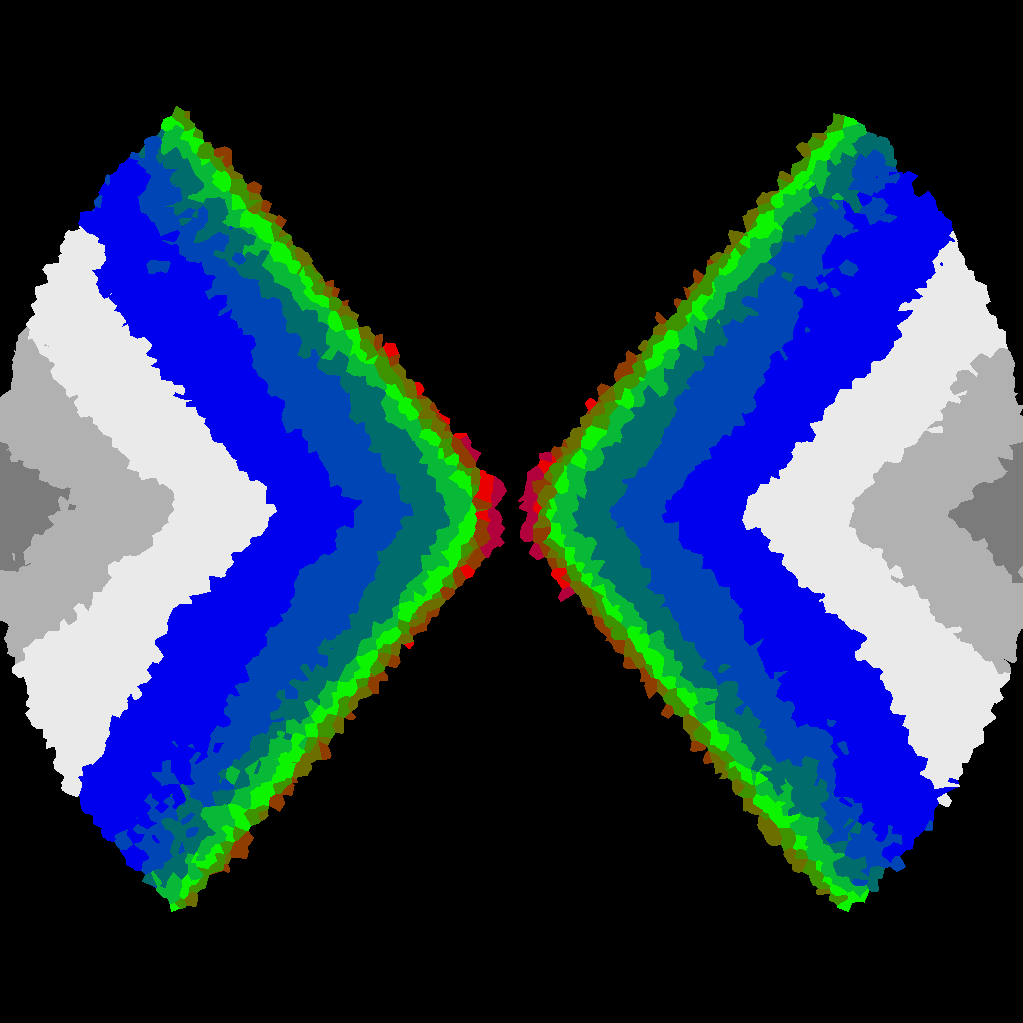} \put(30,93){\color{white} $T_\mathrm{dust}$ \small -- Voronoi grid} \end{overpic}}%
  \fbox{\begin{overpic}[height=0.32\textwidth]{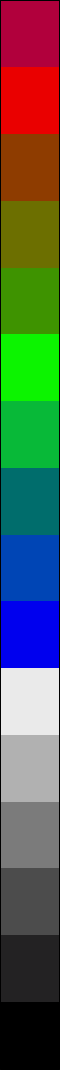} \end{overpic}}\\
  \fbox{\begin{overpic}[width=0.32\textwidth]{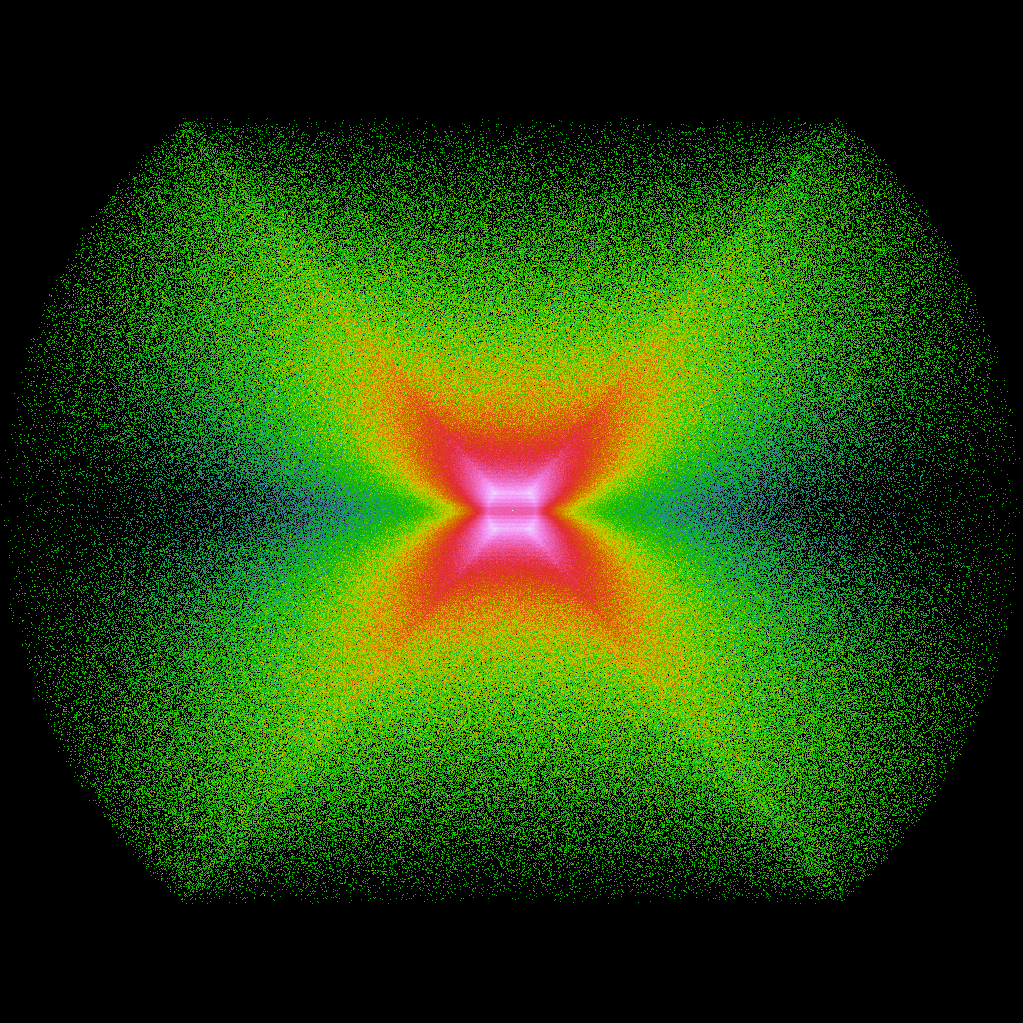} \put(28,93){\color{white} $f$ \small -- cylindrical grid} \end{overpic}}%
  \fbox{\begin{overpic}[width=0.32\textwidth]{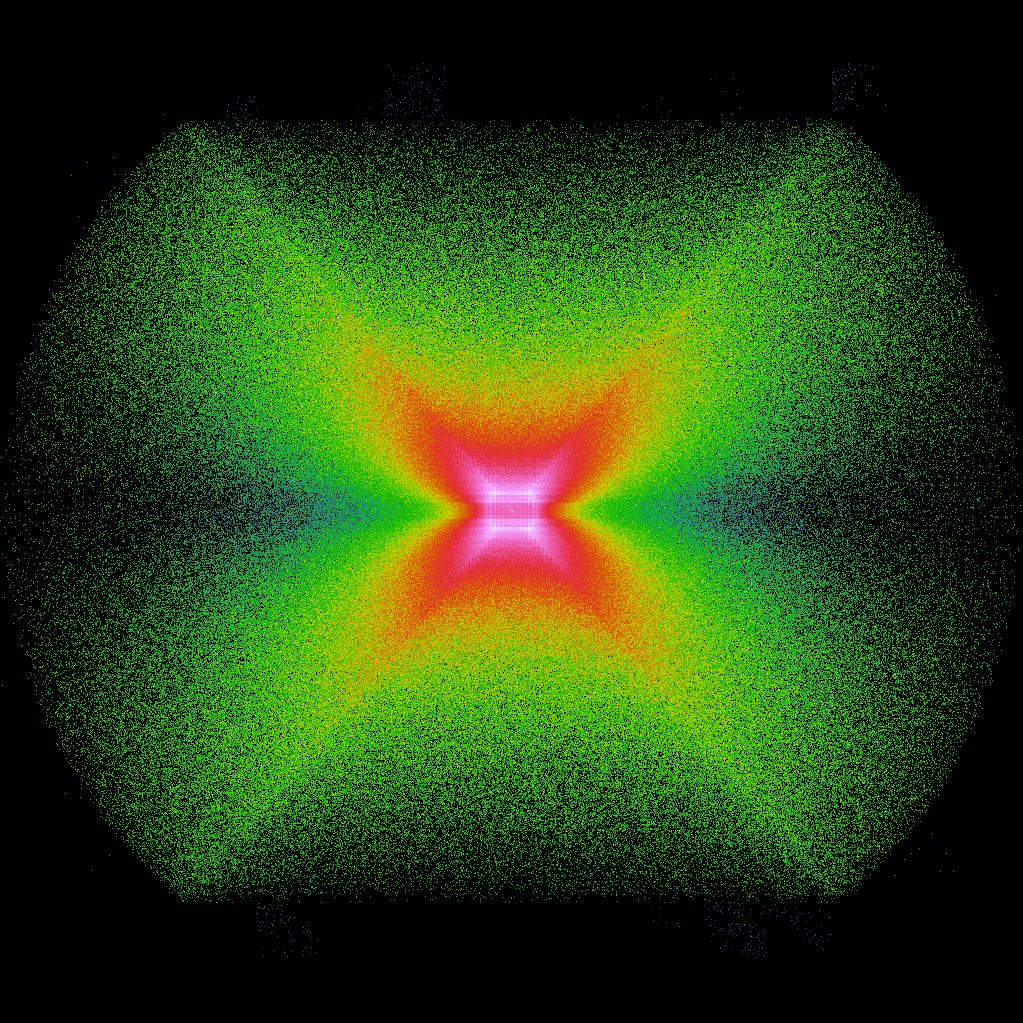} \put(32,93){\color{white} $f$ \small -- octree grid} \end{overpic}}%
  \fbox{\begin{overpic}[width=0.32\textwidth]{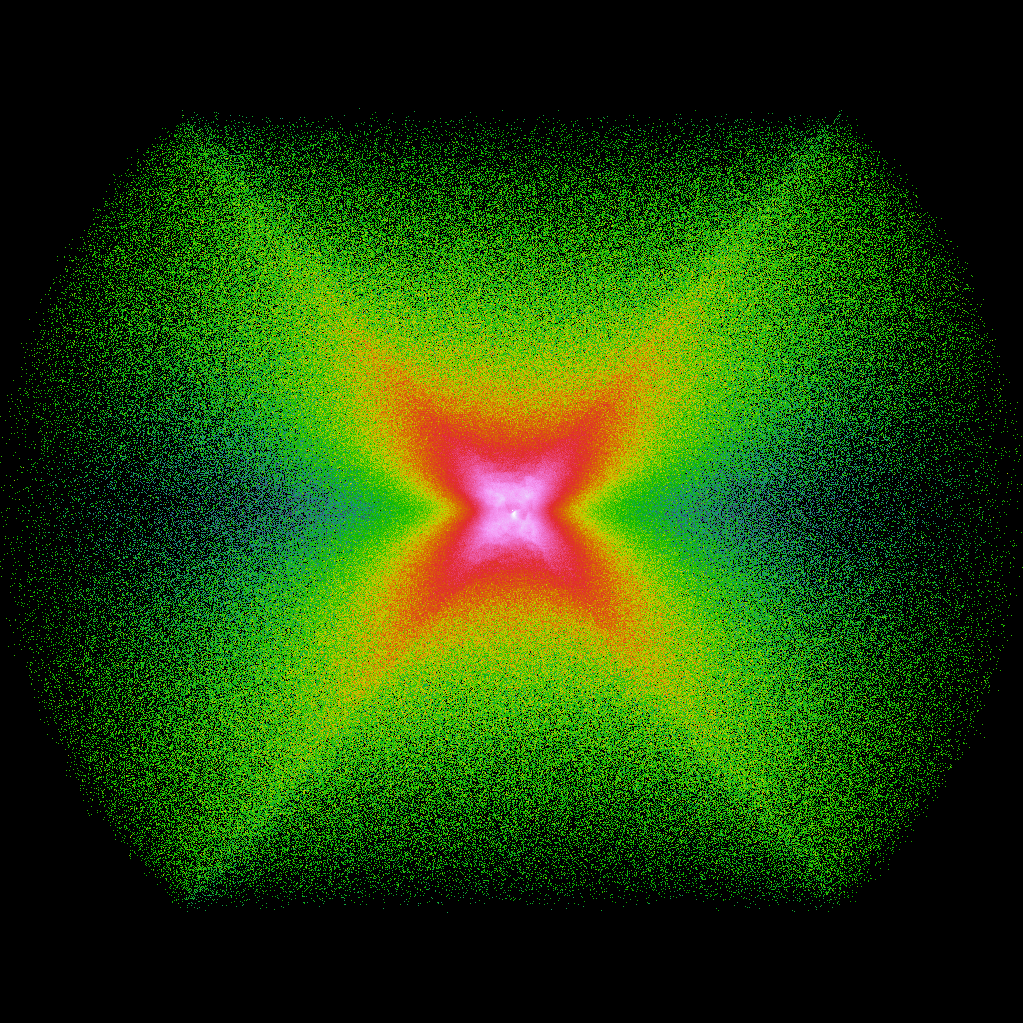} \put(30,93){\color{white} $f$ \small -- Voronoi grid} \end{overpic}}%
  \fbox{\begin{overpic}[height=0.32\textwidth]{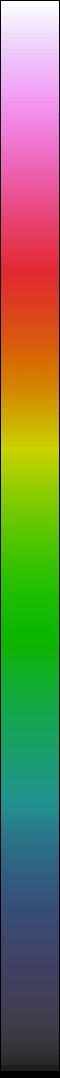} \end{overpic}}
  \caption{Illustration of the results for the \emph{torus} model with three different dust grids.
              \textbf{Rows} -- 
              \emph{top}: the dust density distribution (cut through the central edge-on plane);
              \emph{middle}: the calculated dust temperature (cut through the central edge-on plane);
              \emph{bottom}: the calculated flux density escaping from the model (edge-on view).
              \textbf{Columns} -- 
              \emph{left}: regular 2D cylindrical grid with $250^2=62\,500$ cells;
              \emph{middle}: adaptive octree grid with $\approx 950\,000$ cells;
              \emph{right}: Voronoi grid with $\approx 950\,000$ uniformly distributed cells.
              }
  \label{fig:torus}
\end{figure*}

\begin{figure*}
  \centering 
  \setlength\fboxsep{1pt}
  \setlength\fboxrule{0pt}
  \fbox{\begin{overpic}[width=0.485\textwidth,height=0.1\textwidth]{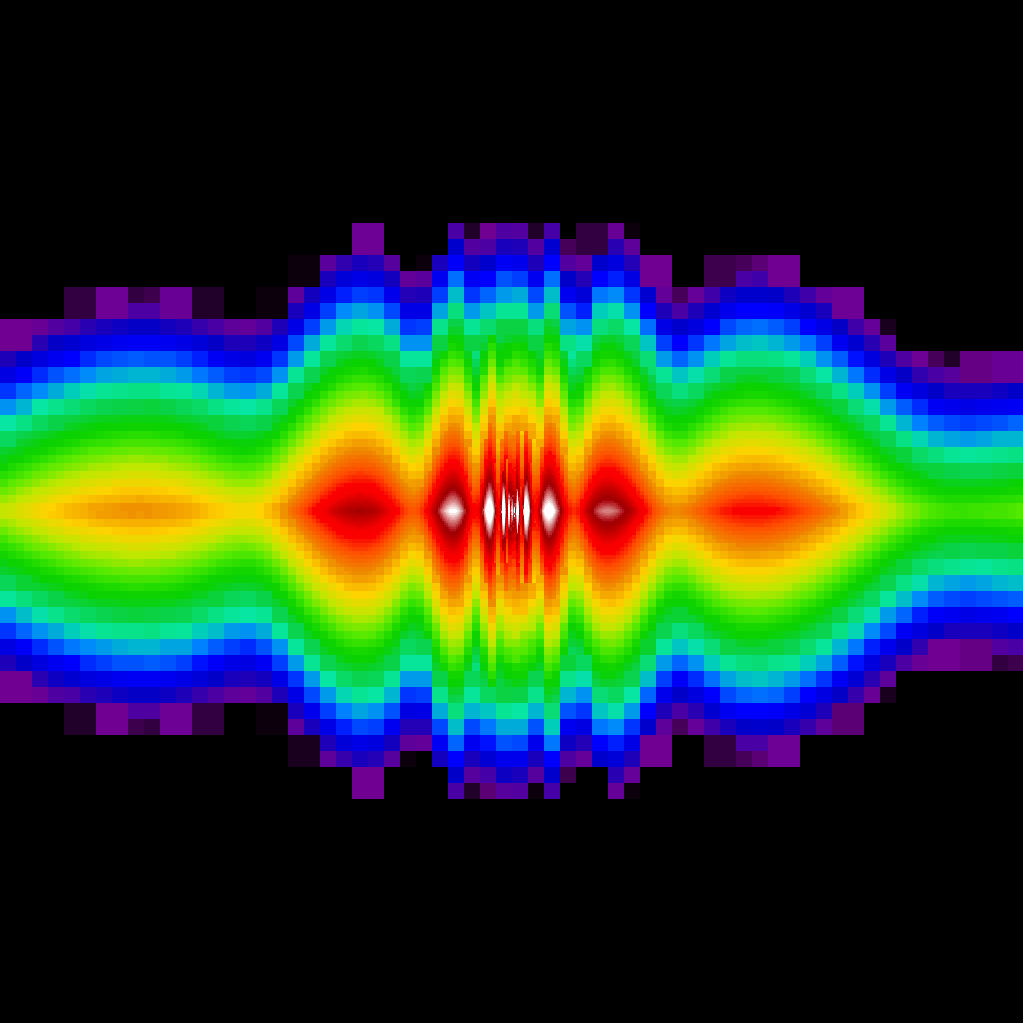} \put(3,17){\color{white} $\rho_\mathrm{dust}$ \small -- octree grid} \end{overpic}}%
  \fbox{\begin{overpic}[width=0.485\textwidth,height=0.1\textwidth]{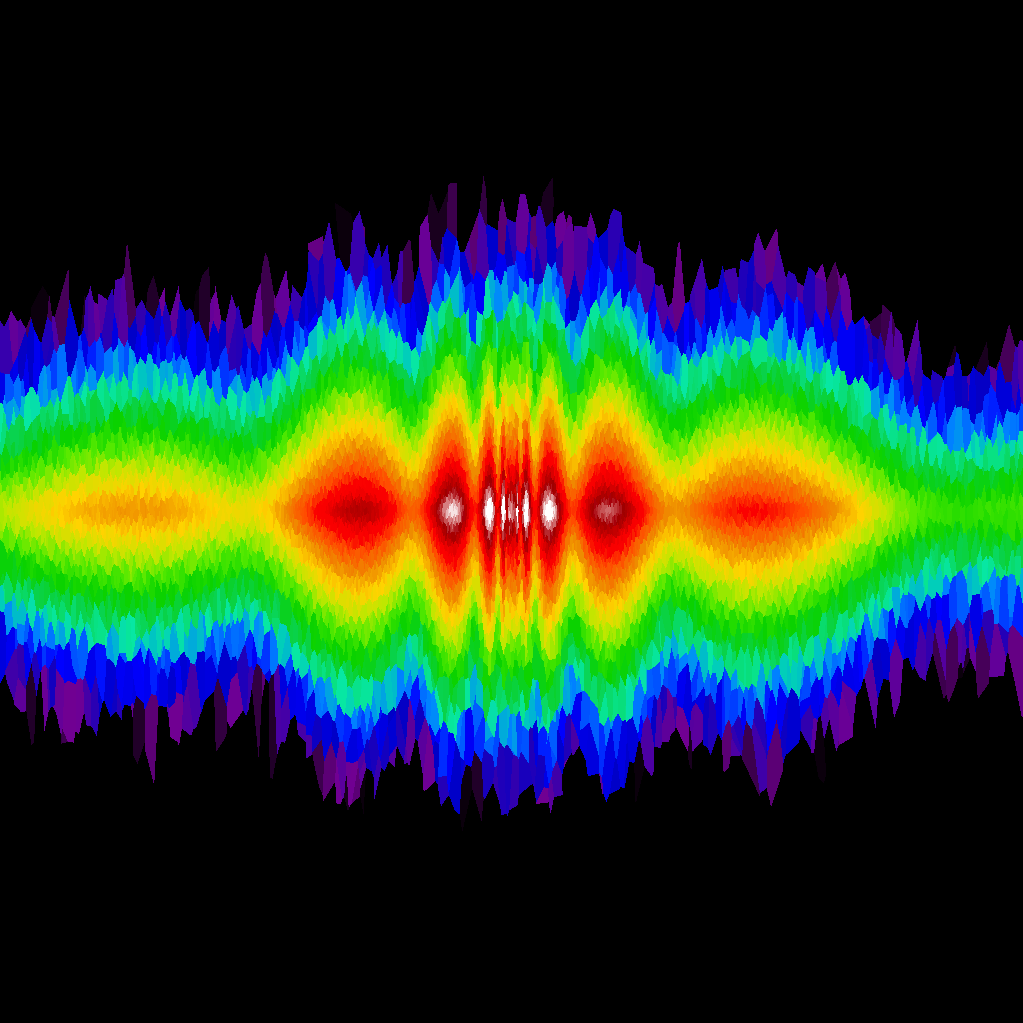} \put(3,17){\color{white} $\rho_\mathrm{dust}$ \small -- Voronoi grid} \end{overpic}}%
  \fbox{\begin{overpic}[width=0.015\textwidth,height=0.1\textwidth]{color_dens.png} \end{overpic}}\\
  \fbox{\begin{overpic}[width=0.485\textwidth,height=0.1\textwidth]{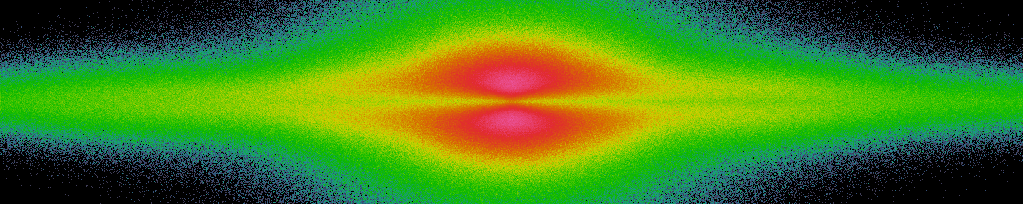} \put(3,17){\color{white} $f$ \small -- octree grid} \end{overpic}}%
  \fbox{\begin{overpic}[width=0.485\textwidth,height=0.1\textwidth]{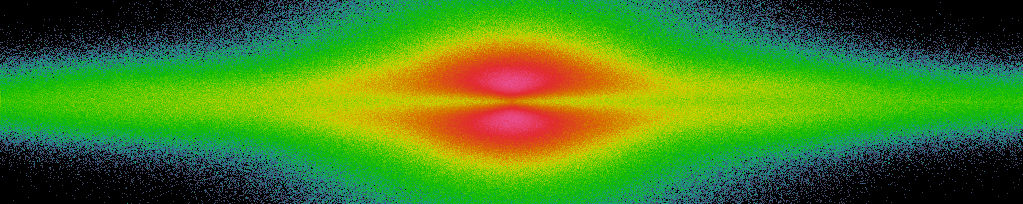} \put(3,17){\color{white} $f$ \small -- Voronoi grid} \end{overpic}}%
  \fbox{\begin{overpic}[width=0.015\textwidth,height=0.1\textwidth]{color_flux.png} \end{overpic}}
  \caption{Illustration of the results for the \emph{spiral} model, edge-on view. 
              \textbf{Rows} -- 
              \emph{top}: the dust density distribution (cut through the central edge-on plane);
              \emph{bottom}: the calculated flux density escaping from the model (edge-on view).
              \textbf{Columns} -- 
              \emph{left}: adaptive octree grid with $\approx 1\,350\,000$ cells;
              \emph{right}: Voronoi grid with $\approx 1\,350\,000$ cells with a non-uniform, weighed distribution.
              }
  \label{fig:spiral_eo}
\end{figure*}

\begin{figure*}%
  \sidecaption%
  \setlength\fboxsep{1pt}%
  \setlength\fboxrule{0pt}%
  \begin{minipage}{0.7\textwidth}%
  \fbox{\begin{overpic}[width=0.485\textwidth]{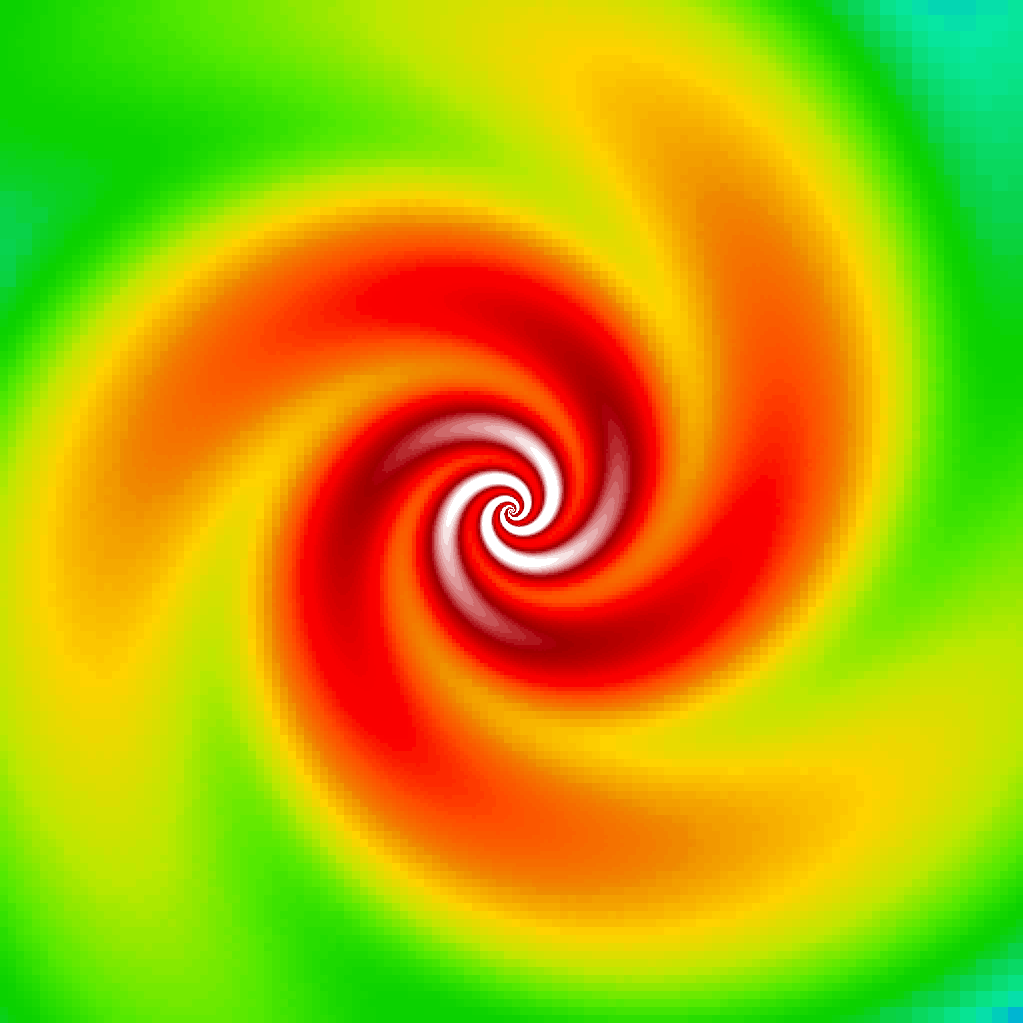} \put(3,93){\color{white} $\rho_\mathrm{dust}$ \small -- octree grid} \end{overpic}}%
  \fbox{\begin{overpic}[width=0.485\textwidth]{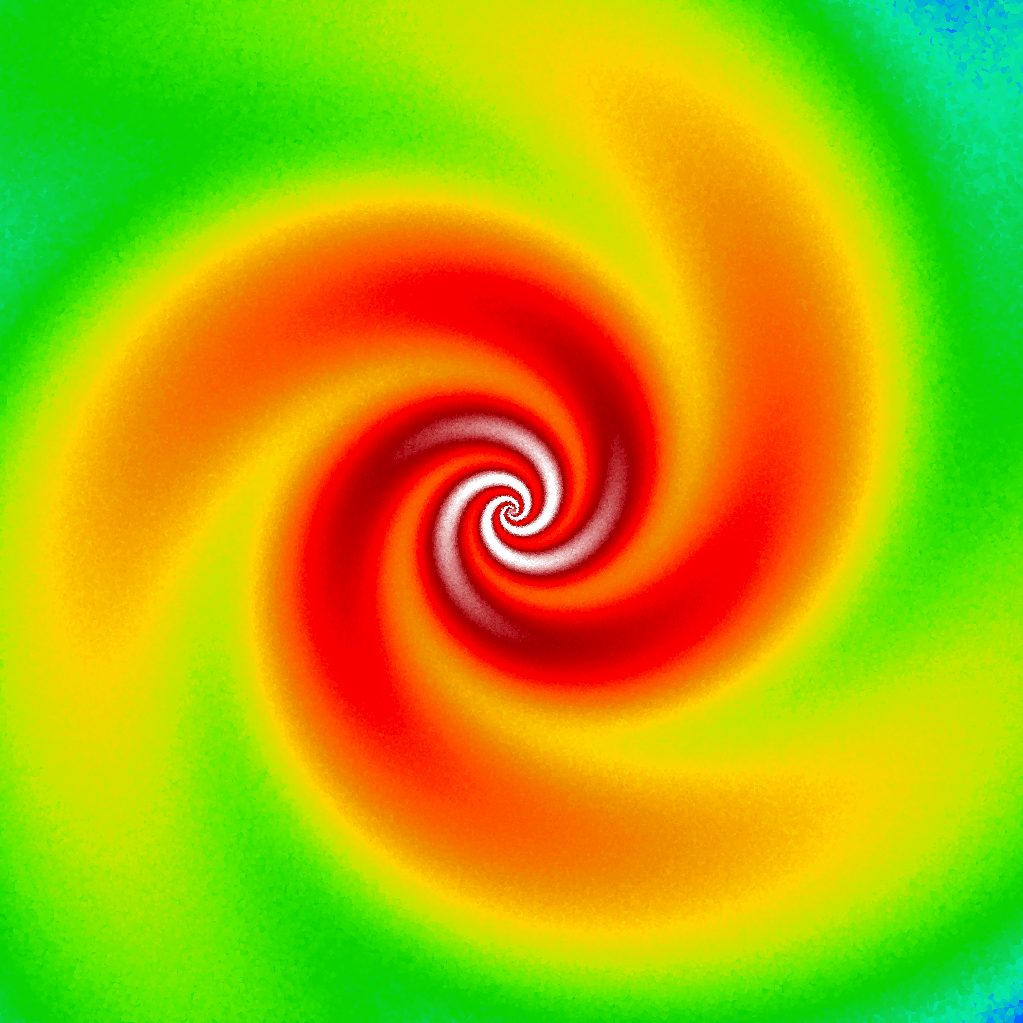} \put(3,93){\color{white} $\rho_\mathrm{dust}$ \small -- Voronoi grid} \end{overpic}}%
  \fbox{\begin{overpic}[height=0.485\textwidth]{color_dens.png} \end{overpic}}\\
  \fbox{\begin{overpic}[width=0.485\textwidth]{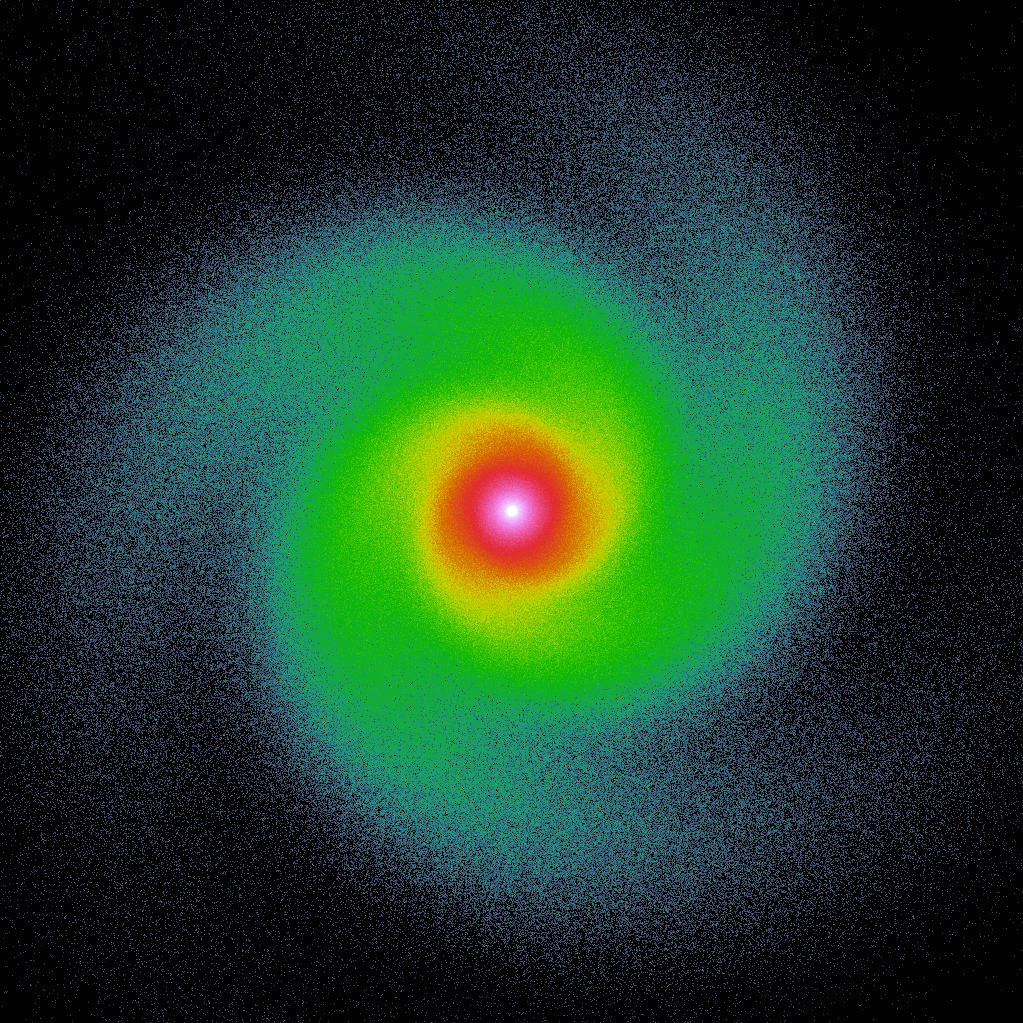} \put(3,93){\color{white} $f$ \small -- octree grid} \end{overpic}}%
  \fbox{\begin{overpic}[width=0.485\textwidth]{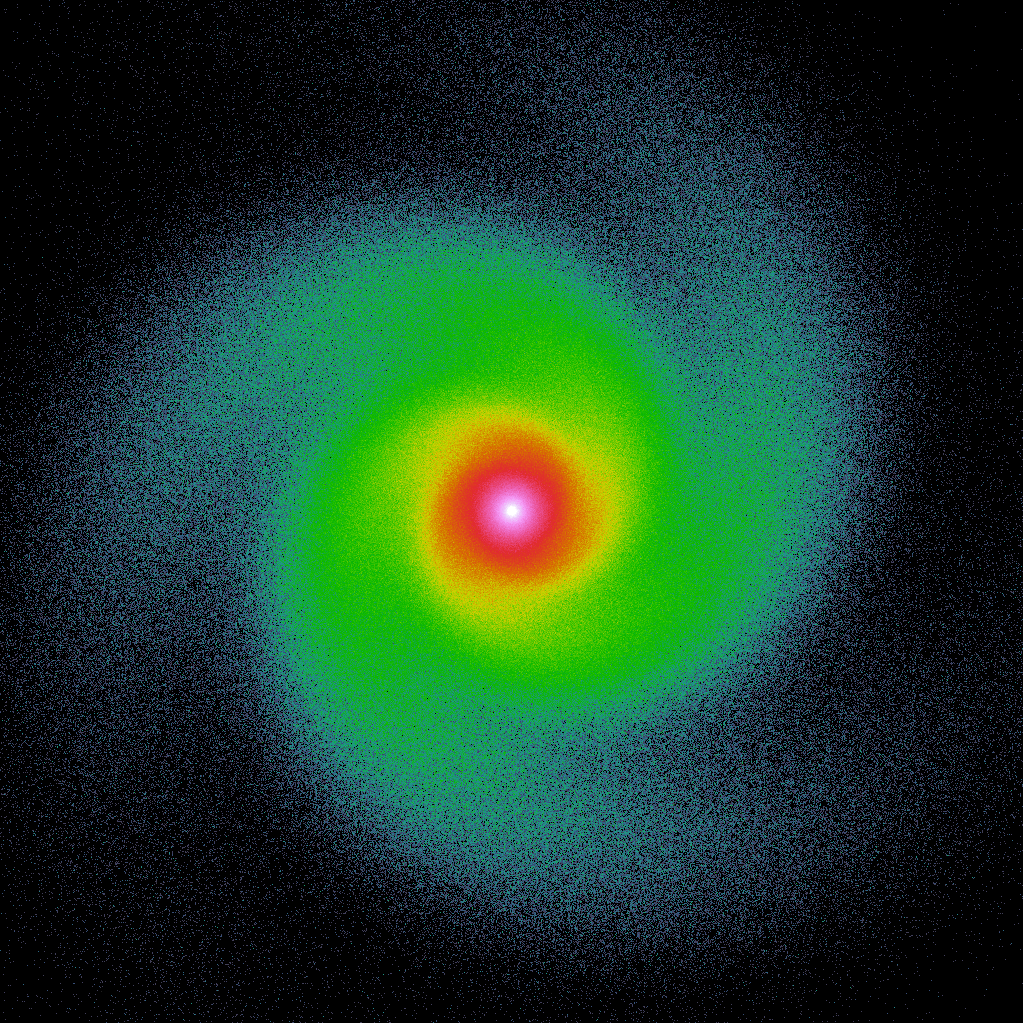} \put(3,93){\color{white} $f$ \small -- Voronoi grid} \end{overpic}}%
  \fbox{\begin{overpic}[height=0.485\textwidth]{color_flux.png} \end{overpic}}%
  \end{minipage}
  \caption{Illustration of the results for the \emph{spiral} model, face-on view. 
              \textbf{Rows} -- 
              \emph{top}: the dust density distribution (cut through the central face-on plane);
              \emph{bottom}: the calculated flux density escaping from the model (face-on view).
              \textbf{Columns} -- 
              \emph{left}: adaptive octree grid with $\approx 1\,350\,000$ cells;
              \emph{right}: Voronoi grid with $\approx 1\,350\,000$ cells with a non-uniform, weighed distribution.
              }
  \label{fig:spiral_fo}
\end{figure*}


\section{Tests, results and discussion}
\label{sec:testing}

\subsection{Implementation}

SKIRT \citep{2011ApJS..196...22B} is a 3D continuum RT code based on the Monte Carlo technique. 
It is used for studying dusty astrophysical objects including
spiral galaxies \citep{2012MNRAS.427.2797D, 2012MNRAS.419..895D, 2012arXiv1212.0538D} 
and active galactic nuclei \cite{2012MNRAS.420.2756S, 2013arXiv1301.4244S}. 
Input models can be defined through a range of built-in geometries, or imported from the results of a MHD simulation.
SKIRT also offers various dust grid options, including regular grids and adaptive grids \citep{2013A&A...554A..10S}.

We implemented a Voronoi dust grid in SKIRT according to the method presented in Sect.\,\ref{sec:method}.
This allowed us to use the built-in geometries for creating synthetic test models, and to compare the results with those produced
by the existing and well-tested grids.

We employed the open source library Voro++ \citep{Voro-lib} to setup the input data described in Sects.\,\ref{sec:straight} and \ref{sec:findcell}.
The library and its data structures are used only during setup. 
All relevant information is extracted and stored in our own data structures for reference after setup.

\subsection{Test models}

We tested the Voronoi dust grid with two synthetic models of our own making, called \emph{torus} and \emph{spiral},
and we ran the RT benchmark described by \citet{2004A&A...417..793P}. We first present the results for our models, 
and in Sect.\,\ref{sec:pascucci} we discuss the results for the Pascucci benchmark.

The \emph{torus} model consists of a central light source surrounded by an axisymmetric dusty torus,
as might be present in the center of active galactic nuclei. 
The dust geometry is described by a radial power-law density from a given inner to outer radius, with an opening angle of $50$ degrees.
A cut through the dust distribution is shown in the top row of Fig.\,\ref{fig:torus}.
The sites for the Voronoi dust grid are selected randomly from a uniform distribution over the cuboidal domain enclosing the torus.
Since the model is axisymmetric, we can compare the results of the Voronoi grid with those produced by a regular two-dimensional (2D) cylindrical grid, 
in addition to those produced by an adaptive (3D) octree grid.
In Fig.\,\ref{fig:resolution} we illustrate the effect of the number of Voronoi grid cells for the torus model.

The \emph{spiral} model represents an idealized spiral galaxy with three arms, similar to the spiral model presented in \citet{2013A&A...554A..10S}.
The stellar distribution includes a flattened S\'ersic bulge and a double-exponential disk with a spiral arm perturbation.
The dust is distributed in a thinner, similarly perturbed double-exponential disk.
Cuts through the dust distribution are shown in the top row of Figs.\,\ref{fig:spiral_eo} and \ref{fig:spiral_fo}.
In this case the sites for the Voronoi dust grid are selected randomly from the dust distribution, as opposed to a uniform distribution.
Areas with a higher dust density are thus -- on average -- covered with smaller cells.


\begin{table}
\caption{\emph{Grid quality.}
             The difference between the theoretical and gridded dust density is sampled at a large number of random points, uniformly distributed over the domain.
             The standard deviation on this difference is used as a quality measure for the grid.
             In the table, the value for the octree grid is normalized to unity for each model.
             \emph{Smaller numbers indicate better quality.}
             }
\label{table:quality}
\centering
\begin{tabular}{c c c c}
\hline\hline
Model & Cylindrical & Octree & Voronoi \\ 
\hline
  Torus & 0.82 & 1 & 1.75 \\
  Spiral & -- & 1 & 1.68 \\
\hline
\end{tabular}
\end{table}

\begin{table}
\caption{\emph{Run time.}
              The elapsed time for the photon shooting phase of a simulation is divided by the number of grid cells crossed during that phase.
	     The result is an indication of the time spent per cell crossing, including grid traversal calculations and some overhead for generating the
	     random paths and for storing results. The tests were performed on a typical desktop computer using a single core.
	     The last column lists the ratio between the run times for the Voronoi and octree grids.
             \emph{Larger numbers indicate slower performance.}
             }
\label{table:timings}
\centering
\begin{tabular}{c c c c c}
\hline\hline
 & & \multicolumn{3}{c}{Time per cell crossing (ns)} \\
Model & Simulation type & Octree & Voronoi & Vor./Oct. \\ 
\hline
  Torus & monochromatic & 219 & 693 & 3.2 \\
  Torus & panchromatic & 400 & 1006 & 2.5 \\
  Spiral & monochromatic & 309 & 903 & 2.9 \\
  Spiral & panchromatic & 442 & 1095 & 2.5 \\
\hline
\end{tabular}
\end{table}


\subsection{Test grids}
\label{sec:grids}

For the torus model we ran simulations with three different dust grids: a regular 2D cylindrical grid with $250^2=62\,500$ cells;
an adaptive octree grid with $\approx 950\,000$ cells; and a Voronoi grid with about the same number of uniformly distributed cells.
The top row of Fig.\,\ref{fig:torus} shows a cut through the gridded dust density distribution for each of these grids.
The cylindrical grid captures the sharp edges of the model perfectly, because the cylindrical coordinate axes are lined up with the edges.
The octree grid does a fine job as well due to its adaptive nature: smaller cells are automatically created along the sharp edges.
The Voronoi grid doesn't do particularly well at the edges, due to the random placement of its cells. This would not be an issue
when importing a grid from a moving mesh code, because the cell sizes would already be properly adjusted to the underlying gradients.

For the spiral model we ran simulations with two different dust grids: an adaptive octree grid with $\approx 1\,350\,000$ cells;
and a Voronoi grid with about the same number of cells, placed using a weighed distribution according to the dust density (smaller cells in higher density areas).
The top rows of Figs.\,\ref{fig:spiral_eo} and \ref{fig:spiral_fo} show a cut through the gridded dust density distribution for each of these grids.
The differences between the grids are most easily seen in the lower density areas.

Although this study does not focus on grid quality, we still need to ensure that our Voronoi grid implementation properly represents the 
theoretical dust densities defined by the synthetic models.
To obtain an objective quality measure, we sample the theoretical dust density $\rho_\textrm{t}$ and the gridded dust density $\rho_\textrm{g}$
at a large number of random points uniformly distributed over the domain. 
We use the standard deviation of the difference $\rho_\textrm{t}-\rho_\textrm{g}$ as a measure for how well the grid reflects the theoretical density distribution.
Table \ref{table:quality} lists the resulting numbers for the various grids and models.
For each model the value for the octree grid is normalized to unity.

Taking into account our naive cell placement, the Voronoi grid compares well with the highly tuned adaptive octree grid,
thus verifying this aspect of our implementation.

\subsection{Results}

Shooting photon packages through the grid is the more important test in the context of this study.

The middle row of Fig.\,\ref{fig:torus} shows the dust temperature calculated by
a panchromatic simulation for the torus model, using the three grids describe above.
All quantities, including the radiation field and the amount of dust absorption, are discretized on the same grid as the dust density.
In each simulation, the central light source emits $10^5$ photon packages for each of 100 wavelength bins on a logarithmic grid.
Scattering events cause additional photon packages to be created, which is
particularly relevant for this model due to the high optical depth of the torus.
In the end each simulation traces about 700 million photon packages through the dust grid.

The bottom rows of Figs.\,\ref{fig:torus}, \ref{fig:spiral_eo} and \ref{fig:spiral_fo} show the flux density calculated by
a monochromatic simulation for each model and grid combination.
The Poisson noise is caused by the statistical nature of the Monte Carlo technique.
In each simulation, the light sources emit 10 million photon packages at a fixed wavelength, and
scattering events again cause additional photon packages to be created.

Other than the effects of grid resolution and the unavoidable noise, the calculated temperature and flux density maps are the same for the various grids.
In particular, as noted in Sect.\,\ref{sec:grids}, the Voronoi grid does not resolve the central area of the dust distribution as well as the other grids,
causing some deviation in the central area of the calculated flux density field. 
This effect is ultimately due to the naive placement of the Voronoi cells in our tests, and would not be present for a properly adjusted grid.

These results validate the accuracy of our straight path calculation method for Voronoi grids.

Table \ref{table:timings} provides an indication of the processing time spent per cell crossing for each simulation. To obtain these numbers, 
the elapsed time for the photon shooting phase of a simulation is divided by the number of grid cells crossed during that phase.
The result thus includes some overhead for generating the random paths and for storing results, in addition to the grid traversal calculation itself.
The tests were performed on a typical desktop computer using a single core.
The last column lists the ratio between the cell crossing times for the Voronoi and octree grids.
The Voronoi grid performs roughly three times slower than our highly optimized
octree implementation (which maintains, for example, a neighbor list for each cell to accelerate the process of finding the next cell on a path).
This seems surprisingly fast in view of the high geometric complexity of a Voronoi grid
(illustrated in Figs.\,\ref{fig:tessellation} and \ref{fig:cell}) compared to the cuboidal cells in an octree.
Moreover, as noted in the introduction, an octree grid may need many more cells than the Voronoi grid to represent a particular density field,
further balancing performance in favor of the Voronoi grid.

As discussed in Sect.\,\ref{sec:straight}, the cell crossing algorithm may occasionally fail to find an exit point due to computational inaccuracies.
In our tests this occurred at most once per 50 million cell crossings, so this issue does not affect the algorithm's performance.

\subsection{The Pascucci benchmark}
\label{sec:pascucci}

\begin{figure*}
  \centering 
  \setlength\fboxsep{1pt}
  \setlength\fboxrule{0pt}
  \raisebox{0.03\textwidth}{
  \fbox{\begin{overpic}[width=0.33\textwidth,height=0.30\textwidth]{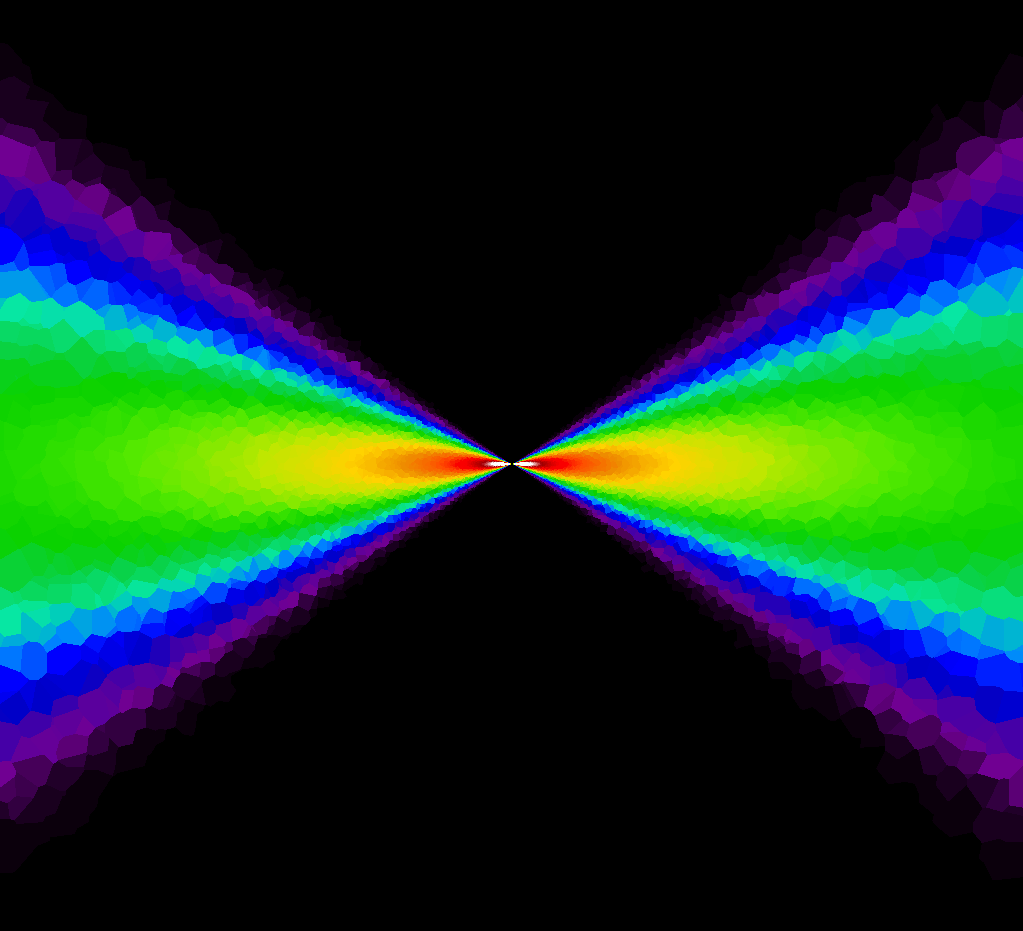} \put(32,83){\color{white} $\rho_\mathrm{dust}$ \small -- Voronoi grid} \end{overpic}}%
  }%
  \fbox{\begin{overpic}[width=0.33\textwidth,height=0.33\textwidth]{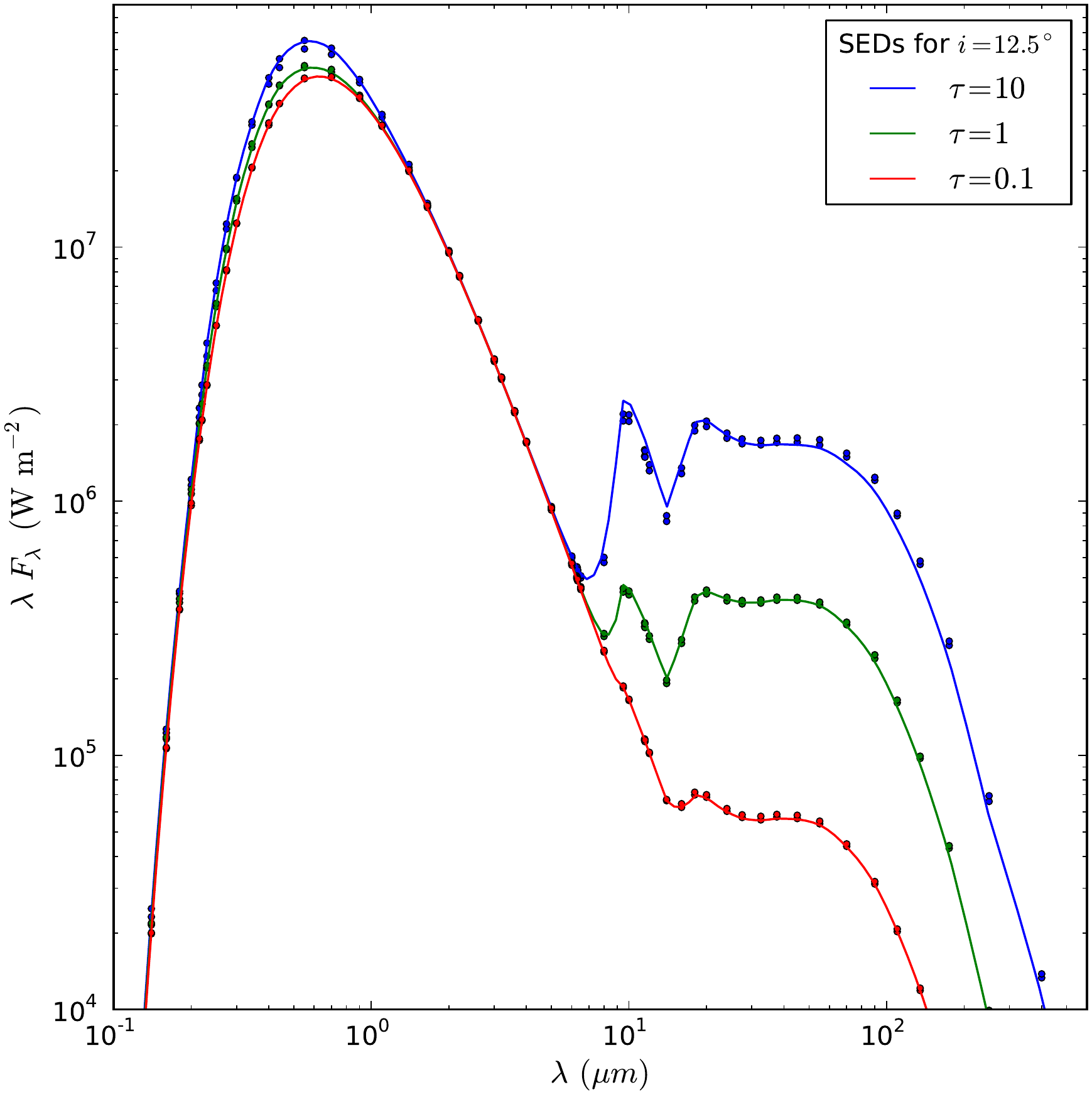} \end{overpic}}%
  \fbox{\begin{overpic}[width=0.33\textwidth,height=0.33\textwidth]{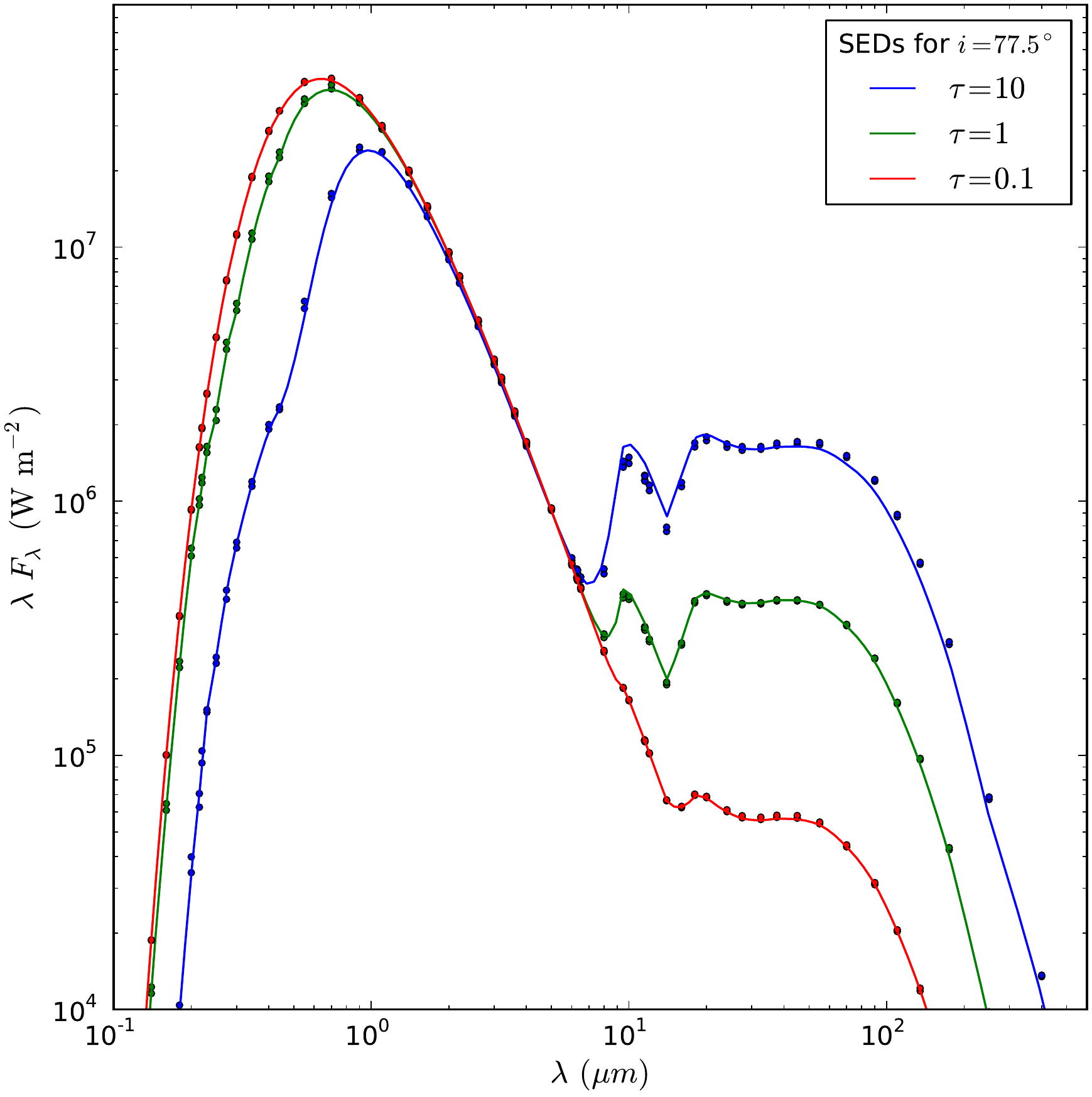} \end{overpic}}
  \caption{Illustration of the results for the \emph{Pascucci} benchmark \citep{2004A&A...417..793P}.
  	       The \emph{left} panel shows a cut through the central edge-on plane of the dust density distribution discretized on a 3D Voronoi grid 
	       with one million cells randomly placed according to a $1/r$ distribution.
	       The other panels show the simulated spectral energy distribution (SED) 
	       for disk inclinations equal to 12.5$^\circ$ (\emph{center}) and 77.5$^\circ$ (\emph{right}), for optical depths $\tau=0.1,1$ and $10$.
	       Dots indicate benchmark reference points; solid lines represent our simulation results
              using the 3D Voronoi grid shown in the left panel.}
  \label{fig:pascucci}
\end{figure*}

The \emph{Pascucci} benchmark \citep{2004A&A...417..793P} models a star embedded in a circumstellar disk with an inner cavity free of dust,
prescribing an analytical 2D distribution and a set of optical depths and viewing angles.
A cut through the central edge-on plane of the dust density distribution is shown in the left panel of Fig.\,\ref{fig:pascucci}.

We ran panchromatic simulations for this model with optical depths $\tau=0.1,1$ and $10$
using a 3D Voronoi grid consisting of one million cells randomly placed according to a $1/r$ distribution.  
This distribution serves to properly resolve the intense radiation field in the center of the model.
In each simulation, the central light source emits $10^5$ photon packages for each of 150 wavelength bins on a logarithmic grid.

The two rightmost panels of Fig.\,\ref{fig:pascucci} compare the spectral energy distribution (SED) produced by our SKIRT simulations
with the corresponding benchmark results published in \citet{2004A&A...417..793P}.
The center panel shows the SEDs for the various optical depths at a nearly face-on disk inclination of 12.5$^\circ$,
the right panel at a nearly edge-on inclination of 77.5$^\circ$.
Dots indicate benchmark reference points; solid lines represent our simulation results.

For higher optical depths our simulation results deviate slightly because the SKIRT code is not optimized for operation in this regime;
running the benchmark with a 2D axially symmetric logarithmic grid results in the same deviation (not shown). These results
further validate our method for calculating straight paths through a Voronoi grid.

\subsection{Applicability}

From Sect.\,\ref{sec:method} it follows that the presented method requires as input data solely the coordinates of the Voronoi sites,
plus any relevant physical properties (such as mass densities) for the cell surrounding each site.
In other words the interface between the input model and the RT code is very thin, opening up a wide range of possibilities.
An input model can be defined by SPH particles, serving as Voronoi sites; or by a Voronoi mesh produced by an MHD code;
or by appropriately distributed random points generated from (semi)-analytical density or opacity fields, similar to
the approach in e.g.\ \citet{2010A&A...515A..79P}.

We also note in Sects.\,\ref{sec:straight} and \ref{sec:findcell} that the path calculation algorithm itself requires no information on a Voronoi cell
other than its bounding box and the locations of its own site and all neighboring sites.
The required data structures can be easily built from the input data using a publicly available Voronoi library. 
The library code is invoked only during the initialization phase, minimizing its impact on performance and robustness,
and allowing it to be easily replaced by another code if the need arises.
For example, while we are happy with the Voro++ library's ease of use and with its performance during the tests,
we may in the future consider using a parallelized method \citep{Lo201288, 2011arXiv1109.2218S}.

As a consequence the presented method allows the RT code to support a Voronoi grid while remaining uncoupled from the code
producing the input model. This decreases the complexity of the interface, and allows cooperation even 
when source code is not publicly available. In contrast to this approach, for example, the Sunrise RT code \citep{2006MNRAS.372....2J} 
directly invokes parts of the non-public Arepo moving mesh code \citep{2011arXiv1109.2218S} to implement the interface 
and to build its Voronoi grid \citep{Sunrise-Arepo}.


\section{Conclusions}
\label{sec:conclusions}

The choice of an appropriate discretization is crucial in any numerical simulation code. Due to the large dynamic range of the 
physical quantities, in most problems the resolution of the grid must scale with the field densities or gradients.
Adaptive grids with cuboidal cells, such as octrees or more generally AMR grids, have proven very popular in part because
of their relative ease of implementation.
However several recent codes have adopted unstructured grids based on Voronoi tessellations, or equivalently, Delaunay triangulations.
These grids tend to more closely reflect dynamic ranges in the model with fewer cells,
presenting cell boundaries that are more adjusted to the underlying gradients. 
Since a Voronoi grid is defined solely by its generating points, the cell size and distribution can be easily fine-tuned by placing these
sites in the appropriate locations.

In a RT simulation the Voronoi grid can be a very flexible tool. 
Appropriate sites can be generated randomly, distributed according to the input model's density or opacity fields; if needed extra sites
can be added in high-gradient areas.
In the case of a particle-based input model, the particle locations themselves can serve a sites; and for an input model already
based on a Voronoi mesh no re-gridding is required at all.

In this work we have shown that it is straightforward to implement accurate and efficient RT on Voronoi grids.
In spite of the geometric complexity of the cell boundaries, calculating straight paths between two arbitrary points
through a 3D Voronoi grid is only about three times slower than a highly optimized octree implementation with the same number of cells,
while in practice the total number of Voronoi grid cells may be lower for an equally good representation of the density field.
The presented method automatically yields the precise distance covered by the path inside each grid cell, and eliminates the need
for approximate corrections or work-arounds required by alternate approaches where the radiation travels only along the Delaunay edges. 
The method requires only a thin interface with the input model and with the actual construction of the grid,
allowing codes to remain largely uncoupled and enabling the use of a publicly available Voronoi library.

While we implemented and tested the method in our continuum RT code SKIRT, focusing on the effects of dust, it is widely applicable 
to all RT codes using ray tracing or Monte Carlo techniques. 

We conclude that the benefits of using a Voronoi grid in RT simulation codes will often 
outweigh the somewhat slower performance.


\begin{acknowledgements}
This work fits in the CHARM framework (Contemporary physical challenges in Heliospheric and AstRophysical Models),
a phase VII Interuniversity Attraction Pole (IAP) programme organised by BELSPO, the BELgian federal Science Policy Office.
WS acknowledges the support of Al-Baath University and The Ministry of High Education in Syria in the form of a research grant. 
\end{acknowledgements}


\bibliographystyle{aa} 
\bibliography{voro}

\begin{thebibliography}{93}
\expandafter\ifx\csname natexlab\endcsname\relax\def\natexlab#1{#1}\fi

\bibitem[{{Abdikamalov} {et~al.}(2012){Abdikamalov}, {Burrows}, {Ott},
  {L{\"o}ffler}, {O'Connor}, {Dolence}, \& {Schnetter}}]{2012ApJ...755..111A}
{Abdikamalov}, E., {Burrows}, A., {Ott}, C.~D., {et~al.} 2012, \apj, 755, 111

\bibitem[{{Acreman} {et~al.}(2010){Acreman}, {Harries}, \&
  {Rundle}}]{2010MNRAS.403.1143A}
{Acreman}, D.~M., {Harries}, T.~J., \& {Rundle}, D.~A. 2010, \mnras, 403, 1143

\bibitem[{{Baes} {et~al.}(2003){Baes}, {Davies}, {Dejonghe}, {Sabatini},
  {Roberts}, {Evans}, {Linder}, {Smith}, \& {de Blok}}]{2003MNRAS.343.1081B}
{Baes}, M., {Davies}, J.~I., {Dejonghe}, H., {et~al.} 2003, \mnras, 343, 1081

\bibitem[{{Baes} \& {Dejonghe}(2001)}]{2001MNRAS.326..733B}
{Baes}, M. \& {Dejonghe}, H. 2001, \mnras, 326, 733

\bibitem[{{Baes} \& {Dejonghe}(2002)}]{2002MNRAS.335..441B}
{Baes}, M. \& {Dejonghe}, H. 2002, \mnras, 335, 441

\bibitem[{{Baes} {et~al.}(2011){Baes}, {Verstappen}, {De Looze}, {Fritz},
  {Saftly}, {Vidal P{\'e}rez}, {Stalevski}, \& {Valcke}}]{2011ApJS..196...22B}
{Baes}, M., {Verstappen}, J., {De Looze}, I., {et~al.} 2011, \apjs, 196, 22

\bibitem[{{Bauer} \& {Springel}(2012)}]{2012MNRAS.423.2558B}
{Bauer}, A. \& {Springel}, V. 2012, \mnras, 423, 2558

\bibitem[{{Bethell} {et~al.}(2004){Bethell}, {Zweibel}, {Heitsch}, \&
  {Mathis}}]{2004ApJ...610..801B}
{Bethell}, T.~J., {Zweibel}, E.~G., {Heitsch}, F., \& {Mathis}, J.~S. 2004,
  \apj, 610, 801

\bibitem[{{Bianchi}(2008)}]{2008A&A...490..461B}
{Bianchi}, S. 2008, \aap, 490, 461

\bibitem[{{Bianchi} {et~al.}(2000){Bianchi}, {Ferrara}, {Davies}, \&
  {Alton}}]{2000MNRAS.311..601B}
{Bianchi}, S., {Ferrara}, A., {Davies}, J.~I., \& {Alton}, P.~B. 2000, \mnras,
  311, 601

\bibitem[{{Brinch} \& {Hogerheijde}(2010)}]{2010A&A...523A..25B}
{Brinch}, C. \& {Hogerheijde}, M.~R. 2010, \aap, 523, A25

\bibitem[{{Chakrabarti} {et~al.}(2007){Chakrabarti}, {Cox}, {Hernquist},
  {Hopkins}, {Robertson}, \& {Di Matteo}}]{2007ApJ...658..840C}
{Chakrabarti}, S., {Cox}, T.~J., {Hernquist}, L., {et~al.} 2007, \apj, 658, 840

\bibitem[{{Ciardi} {et~al.}(2001){Ciardi}, {Ferrara}, {Marri}, \&
  {Raimondo}}]{2001MNRAS.324..381C}
{Ciardi}, B., {Ferrara}, A., {Marri}, S., \& {Raimondo}, G. 2001, \mnras, 324,
  381

\bibitem[{{Code} \& {Whitney}(1995)}]{1995ApJ...441..400C}
{Code}, A.~D. \& {Whitney}, B.~A. 1995, \apj, 441, 400

\bibitem[{{Collins} {et~al.}(2010){Collins}, {Xu}, {Norman}, {Li}, \&
  {Li}}]{2010ApJS..186..308C}
{Collins}, D.~C., {Xu}, H., {Norman}, M.~L., {Li}, H., \& {Li}, S. 2010, \apjs,
  186, 308

\bibitem[{{De Geyter} {et~al.}(2012){De Geyter}, {Baes}, {Fritz}, \&
  {Camps}}]{2012arXiv1212.0538D}
{De Geyter}, G., {Baes}, M., {Fritz}, J., \& {Camps}, P. 2012, ArXiv e-prints

\bibitem[{{De Looze} {et~al.}(2012{\natexlab{a}}){De Looze}, {Baes}, {Bendo},
  {Ciesla}, {Cortese}, {de Geyter}, {Groves}, {Boquien}, {Boselli}, {Brondeel},
  {Cooray}, {Eales}, {Fritz}, {Galliano}, {Gentile}, {Gordon}, {Hony}, {Law},
  {Madden}, {Sauvage}, {Smith}, {Spinoglio}, \&
  {Verstappen}}]{2012MNRAS.427.2797D}
{De Looze}, I., {Baes}, M., {Bendo}, G.~J., {et~al.} 2012{\natexlab{a}},
  \mnras, 427, 2797

\bibitem[{{De Looze} {et~al.}(2012{\natexlab{b}}){De Looze}, {Baes}, {Fritz},
  \& {Verstappen}}]{2012MNRAS.419..895D}
{De Looze}, I., {Baes}, M., {Fritz}, J., \& {Verstappen}, J.
  2012{\natexlab{b}}, \mnras, 419, 895

\bibitem[{{Decin} {et~al.}(2012){Decin}, {Cox}, {Royer}, {Van Marle},
  {Vandenbussche}, {Ladjal}, {Kerschbaum}, {Ottensamer}, {Barlow}, {Blommaert},
  {Gomez}, {Groenewegen}, {Lim}, {Swinyard}, {Waelkens}, \&
  {Tielens}}]{2012A&A...548A.113D}
{Decin}, L., {Cox}, N.~L.~J., {Royer}, P., {et~al.} 2012, \aap, 548, A113

\bibitem[{Delaunay(1934)}]{1934CSMN.7.793}
Delaunay, B. 1934, Classe des Sciences Math{\'e}matiques et Naturelles, 7, 793

\bibitem[{Dirichlet(1850)}]{crll.1850.40.209}
Dirichlet, L. 1850, Journal f{\"u}r die reine und angewandte Mathematik, 40,
  209

\bibitem[{{Disney} {et~al.}(1989){Disney}, {Davies}, \&
  {Phillipps}}]{1989MNRAS.239..939D}
{Disney}, M., {Davies}, J., \& {Phillipps}, S. 1989, \mnras, 239, 939

\bibitem[{{Dolag} \& {Stasyszyn}(2009)}]{2009MNRAS.398.1678D}
{Dolag}, K. \& {Stasyszyn}, F. 2009, \mnras, 398, 1678

\bibitem[{{Doty} {et~al.}(2005){Doty}, {Metzler}, \&
  {Palotti}}]{2005MNRAS.362..737D}
{Doty}, S.~D., {Metzler}, R.~A., \& {Palotti}, M.~L. 2005, \mnras, 362, 737

\bibitem[{{Duffell} \& {MacFadyen}(2011)}]{2011ApJS..197...15D}
{Duffell}, P.~C. \& {MacFadyen}, A.~I. 2011, \apjs, 197, 15

\bibitem[{{Ercolano} {et~al.}(2005){Ercolano}, {Barlow}, \&
  {Storey}}]{2005MNRAS.362.1038E}
{Ercolano}, B., {Barlow}, M.~J., \& {Storey}, P.~J. 2005, \mnras, 362, 1038

\bibitem[{{Fallscheer} {et~al.}(2013){Fallscheer}, {Reid}, {Di Francesco},
  {Martin}, {Hennemann}, {Hill}, {Nguyen-Luong}, {Motte}, {Men'shchikov},
  {Andre}, {Ward-Thompson}, {Griffin}, {Kirk}, {Konyves}, {Rygl}, {Sauvage},
  {Schneider}, {Anderson}, {Benedettini}, {Bernard}, {Bontemps}, {Ginsburg},
  {Molinari}, {Polychroni}, {Rivera-Ingraham}, {Roussel}, {Testi}, {White},
  {Williams}, {Wilson}, {Wong}, \& {Zavagno}}]{2013arXiv1307.0022F}
{Fallscheer}, C., {Reid}, M.~A., {Di Francesco}, J., {et~al.} 2013, ArXiv
  e-prints

\bibitem[{Friedman {et~al.}(1977)Friedman, Bentley, \&
  Finkel}]{Friedman:1977:AFB:355744.355745}
Friedman, J.~H., Bentley, J.~L., \& Finkel, R.~A. 1977, ACM Trans. Math.
  Softw., 3, 209

\bibitem[{{Fritz} {et~al.}(2012){Fritz}, {Gentile}, {Smith}, {Gear}, {Braun},
  {Duval}, {Bendo}, {Baes}, {Eales}, {Verstappen}, {Blommaert}, {Boquien},
  {Boselli}, {Clements}, {Cooray}, {Cortese}, {De Looze}, {Ford}, {Galliano},
  {Gomez}, {Gordon}, {Lebouteiller}, {O'Halloran}, {Kirk}, {Madden}, {Page},
  {Remy}, {Roussel}, {Spinoglio}, {Thilker}, {Vaccari}, {Wilson}, \&
  {Waelkens}}]{2012A&A...546A..34F}
{Fritz}, J., {Gentile}, G., {Smith}, M.~W.~L., {et~al.} 2012, \aap, 546, A34

\bibitem[{{Fromang} {et~al.}(2006){Fromang}, {Hennebelle}, \&
  {Teyssier}}]{2006A&A...457..371F}
{Fromang}, S., {Hennebelle}, P., \& {Teyssier}, R. 2006, \aap, 457, 371

\bibitem[{{Goldsmith} {et~al.}(2008){Goldsmith}, {Heyer}, {Narayanan}, {Snell},
  {Li}, \& {Brunt}}]{2008ApJ...680..428G}
{Goldsmith}, P.~F., {Heyer}, M., {Narayanan}, G., {et~al.} 2008, \apj, 680, 428

\bibitem[{{Goosmann} \& {Gaskell}(2007)}]{2007A&A...465..129G}
{Goosmann}, R.~W. \& {Gaskell}, C.~M. 2007, \aap, 465, 129

\bibitem[{{Gordon} {et~al.}(2001){Gordon}, {Misselt}, {Witt}, \&
  {Clayton}}]{2001ApJ...551..269G}
{Gordon}, K.~D., {Misselt}, K.~A., {Witt}, A.~N., \& {Clayton}, G.~C. 2001,
  \apj, 551, 269

\bibitem[{{Greif} {et~al.}(2011){Greif}, {Springel}, {White}, {Glover},
  {Clark}, {Smith}, {Klessen}, \& {Bromm}}]{2011ApJ...737...75G}
{Greif}, T.~H., {Springel}, V., {White}, S.~D.~M., {et~al.} 2011, \apj, 737, 75

\bibitem[{Guttman(1984)}]{Guttman:1984:RDI:971697.602266}
Guttman, A. 1984, SIGMOD Rec., 14, 47

\bibitem[{{Harries} {et~al.}(2004){Harries}, {Monnier}, {Symington}, \&
  {Kurosawa}}]{2004MNRAS.350..565H}
{Harries}, T.~J., {Monnier}, J.~D., {Symington}, N.~H., \& {Kurosawa}, R. 2004,
  \mnras, 350, 565

\bibitem[{Hayward(2011)}]{Sunrise-Arepo}
Hayward, C. 2011, Running Sunrise with Arepo (wiki page)

\bibitem[{{Hayward} {et~al.}(2011){Hayward}, {Kere{\v s}}, {Jonsson},
  {Narayanan}, {Cox}, \& {Hernquist}}]{2011ApJ...743..159H}
{Hayward}, C.~C., {Kere{\v s}}, D., {Jonsson}, P., {et~al.} 2011, \apj, 743,
  159

\bibitem[{{Heymann} \& {Siebenmorgen}(2012)}]{2012ApJ...751...27H}
{Heymann}, F. \& {Siebenmorgen}, R. 2012, \apj, 751, 27

\bibitem[{{Hubber} {et~al.}(2011){Hubber}, {Batty}, {McLeod}, \&
  {Whitworth}}]{2011A&A...529A..27H}
{Hubber}, D.~A., {Batty}, C.~P., {McLeod}, A., \& {Whitworth}, A.~P. 2011,
  \aap, 529, A27

\bibitem[{{Indebetouw} {et~al.}(2006){Indebetouw}, {Whitney}, {Johnson}, \&
  {Wood}}]{2006ApJ...636..362I}
{Indebetouw}, R., {Whitney}, B.~A., {Johnson}, K.~E., \& {Wood}, K. 2006, \apj,
  636, 362

\bibitem[{{Jonsson}(2006)}]{2006MNRAS.372....2J}
{Jonsson}, P. 2006, \mnras, 372, 2

\bibitem[{{Jonsson} {et~al.}(2010){Jonsson}, {Groves}, \&
  {Cox}}]{2010MNRAS.403...17J}
{Jonsson}, P., {Groves}, B.~A., \& {Cox}, T.~J. 2010, \mnras, 403, 17

\bibitem[{{Juvela} {et~al.}(2012){Juvela}, {Malinen}, \&
  {Lunttila}}]{2012A&A...544A.141J}
{Juvela}, M., {Malinen}, J., \& {Lunttila}, T. 2012, \aap, 544, A141

\bibitem[{{Juvela} \& {Padoan}(2003)}]{2003A&A...397..201J}
{Juvela}, M. \& {Padoan}, P. 2003, \aap, 397, 201

\bibitem[{Keppens {et~al.}(2012)Keppens, Meliani, van Marle, Delmont, Vlasis,
  \& van~der Holst}]{Keppens2012718}
Keppens, R., Meliani, Z., van Marle, A., {et~al.} 2012, Journal of
  Computational Physics, 231, 718

\bibitem[{{Kere{\v s}} {et~al.}(2012){Kere{\v s}}, {Vogelsberger}, {Sijacki},
  {Springel}, \& {Hernquist}}]{2012MNRAS.425.2027K}
{Kere{\v s}}, D., {Vogelsberger}, M., {Sijacki}, D., {Springel}, V., \&
  {Hernquist}, L. 2012, \mnras, 425, 2027

\bibitem[{{Kurosawa} \& {Hillier}(2001)}]{2001A&A...379..336K}
{Kurosawa}, R. \& {Hillier}, D.~J. 2001, \aap, 379, 336

\bibitem[{{Laursen} {et~al.}(2009){Laursen}, {Razoumov}, \&
  {Sommer-Larsen}}]{2009ApJ...696..853L}
{Laursen}, P., {Razoumov}, A.~O., \& {Sommer-Larsen}, J. 2009, \apj, 696, 853

\bibitem[{Lo(2012)}]{Lo201288}
Lo, S. 2012, Computer Methods in Applied Mechanics and Engineering, 237--240,
  88

\bibitem[{{Lunttila} \& {Juvela}(2012)}]{2012A&A...544A..52L}
{Lunttila}, T. \& {Juvela}, M. 2012, \aap, 544, A52

\bibitem[{{Marinacci} {et~al.}(2013){Marinacci}, {Pakmor}, \&
  {Springel}}]{2013arXiv1305.5360M}
{Marinacci}, F., {Pakmor}, R., \& {Springel}, V. 2013, ArXiv e-prints

\bibitem[{{Matthews} \& {Wood}(2001)}]{2001ApJ...548..150M}
{Matthews}, L.~D. \& {Wood}, K. 2001, \apj, 548, 150

\bibitem[{{Misiriotis} {et~al.}(2000){Misiriotis}, {Kylafis}, {Papamastorakis},
  \& {Xilouris}}]{2000A&A...353..117M}
{Misiriotis}, A., {Kylafis}, N.~D., {Papamastorakis}, J., \& {Xilouris}, E.~M.
  2000, \aap, 353, 117

\bibitem[{{Nelson} {et~al.}(2013){Nelson}, {Vogelsberger}, {Genel}, {Sijacki},
  {Kere{\v s}}, {Springel}, \& {Hernquist}}]{2013MNRAS.429.3353N}
{Nelson}, D., {Vogelsberger}, M., {Genel}, S., {et~al.} 2013, \mnras, 429, 3353

\bibitem[{{Niccolini} \& {Alcolea}(2006)}]{2006A&A...456....1N}
{Niccolini}, G. \& {Alcolea}, J. 2006, \aap, 456, 1

\bibitem[{{Paardekooper} {et~al.}(2010){Paardekooper}, {Kruip}, \&
  {Icke}}]{2010A&A...515A..79P}
{Paardekooper}, J.-P., {Kruip}, C.~J.~H., \& {Icke}, V. 2010, \aap, 515, A79

\bibitem[{{Pakmor} {et~al.}(2012){Pakmor}, {Edelmann}, {R{\"o}pke}, \&
  {Hillebrandt}}]{2012MNRAS.424.2222P}
{Pakmor}, R., {Edelmann}, P., {R{\"o}pke}, F.~K., \& {Hillebrandt}, W. 2012,
  \mnras, 424, 2222

\bibitem[{{Paron} {et~al.}(2013){Paron}, {Weidmann}, {Ortega}, {Albacete
  Colombo}, \& {Pichel}}]{2013MNRAS.433.1619P}
{Paron}, S., {Weidmann}, W., {Ortega}, M.~E., {Albacete Colombo}, J.~F., \&
  {Pichel}, A. 2013, \mnras, 433, 1619

\bibitem[{{Pascucci} {et~al.}(2004){Pascucci}, {Wolf}, {Steinacker},
  {Dullemond}, {Henning}, {Niccolini}, {Woitke}, \&
  {Lopez}}]{2004A&A...417..793P}
{Pascucci}, I., {Wolf}, S., {Steinacker}, J., {et~al.} 2004, \aap, 417, 793

\bibitem[{{Pelkonen} {et~al.}(2009){Pelkonen}, {Juvela}, \&
  {Padoan}}]{2009A&A...502..833P}
{Pelkonen}, V.-M., {Juvela}, M., \& {Padoan}, P. 2009, \aap, 502, 833

\bibitem[{{Pinte} {et~al.}(2006){Pinte}, {M{\'e}nard}, {Duch{\^e}ne}, \&
  {Bastien}}]{2006A&A...459..797P}
{Pinte}, C., {M{\'e}nard}, F., {Duch{\^e}ne}, G., \& {Bastien}, P. 2006, \aap,
  459, 797

\bibitem[{{Robitaille}(2011)}]{2011A&A...536A..79R}
{Robitaille}, T.~P. 2011, \aap, 536, A79

\bibitem[{Rycroft(2009)}]{Voro-lib}
Rycroft, C.~H. 2009, Chaos, 19, 041111; doi: 10.1063/1.3215722

\bibitem[{{Saftly} {et~al.}(2013){Saftly}, {Camps}, {Baes}, {Gordon},
  {Vandewoude}, {Rahimi}, \& {Stalevski}}]{2013A&A...554A..10S}
{Saftly}, W., {Camps}, P., {Baes}, M., {et~al.} 2013, \aap, 554, A10

\bibitem[{{Schartmann} {et~al.}(2008){Schartmann}, {Meisenheimer}, {Camenzind},
  {Wolf}, {Tristram}, \& {Henning}}]{2008A&A...482...67S}
{Schartmann}, M., {Meisenheimer}, K., {Camenzind}, M., {et~al.} 2008, \aap,
  482, 67

\bibitem[{{Schechtman-Rook} {et~al.}(2012){Schechtman-Rook}, {Bershady}, \&
  {Wood}}]{2012ApJ...746...70S}
{Schechtman-Rook}, A., {Bershady}, M.~A., \& {Wood}, K. 2012, \apj, 746, 70

\bibitem[{{Sijacki} {et~al.}(2012){Sijacki}, {Vogelsberger}, {Kere{\v s}},
  {Springel}, \& {Hernquist}}]{2012MNRAS.424.2999S}
{Sijacki}, D., {Vogelsberger}, M., {Kere{\v s}}, D., {Springel}, V., \&
  {Hernquist}, L. 2012, \mnras, 424, 2999

\bibitem[{{Springel}(2005)}]{2005MNRAS.364.1105S}
{Springel}, V. 2005, \mnras, 364, 1105

\bibitem[{{Springel}(2010)}]{2010MNRAS.401..791S}
{Springel}, V. 2010, \mnras, 401, 791

\bibitem[{{Springel}(2011)}]{2011arXiv1109.2218S}
{Springel}, V. 2011, ArXiv e-prints

\bibitem[{{Stalevski} {et~al.}(2012){Stalevski}, {Fritz}, {Baes}, {Nakos}, \&
  {Popovi{\'c}}}]{2012MNRAS.420.2756S}
{Stalevski}, M., {Fritz}, J., {Baes}, M., {Nakos}, T., \& {Popovi{\'c}}, L.~{\v
  C}. 2012, \mnras, 420, 2756

\bibitem[{{Stalevski} {et~al.}(2013){Stalevski}, {Fritz}, {Baes}, \&
  {Popovic}}]{2013arXiv1301.4244S}
{Stalevski}, M., {Fritz}, J., {Baes}, M., \& {Popovic}, L.~C. 2013, ArXiv
  e-prints

\bibitem[{{Stamatellos} \& {Whitworth}(2003)}]{2003A&A...407..941S}
{Stamatellos}, D. \& {Whitworth}, A.~P. 2003, \aap, 407, 941

\bibitem[{{Stamatellos} \& {Whitworth}(2005)}]{2005A&A...439..153S}
{Stamatellos}, D. \& {Whitworth}, A.~P. 2005, \aap, 439, 153

\bibitem[{Steinacker {et~al.}(2002)Steinacker, Bacmann, \&
  Henning}]{Steinacker2002765}
Steinacker, J., Bacmann, A., \& Henning, T. 2002, Journal of Quantitative
  Spectroscopy and Radiative Transfer, 75, 765

\bibitem[{{Steinacker} {et~al.}(2006){Steinacker}, {Bacmann}, \&
  {Henning}}]{2006ApJ...645..920S}
{Steinacker}, J., {Bacmann}, A., \& {Henning}, T. 2006, \apj, 645, 920

\bibitem[{{Steinacker} {et~al.}(2005){Steinacker}, {Bacmann}, {Henning},
  {Klessen}, \& {Stickel}}]{2005A&A...434..167S}
{Steinacker}, J., {Bacmann}, A., {Henning}, T., {Klessen}, R., \& {Stickel}, M.
  2005, \aap, 434, 167

\bibitem[{{Steinacker} {et~al.}(2013){Steinacker}, {Baes}, \&
  {Gordon}}]{2013arXiv1303.4998S}
{Steinacker}, J., {Baes}, M., \& {Gordon}, K. 2013, ArXiv e-prints

\bibitem[{{Tasitsiomi}(2006)}]{2006ApJ...645..792T}
{Tasitsiomi}, A. 2006, \apj, 645, 792

\bibitem[{{The Enzo Collaboration} {et~al.}(2013){The Enzo Collaboration},
  {Bryan}, {Norman}, {O'Shea}, {Abel}, {Wise}, {Turk}, {Reynolds}, {Collins},
  {Wang}, {Skillman}, {Smith}, {Harkness}, {Bordner}, {Kim}, {Kuhlen}, {Xu},
  {Goldbaum}, {Hummels}, {Kritsuk}, {Tasker}, {Skory}, {Simpson}, {Hahn},
  {Oishi}, {So}, {Zhao}, {Cen}, \& {Li}}]{2013arXiv1307.2265T}
{The Enzo Collaboration}, {Bryan}, G.~L., {Norman}, M.~L., {et~al.} 2013, ArXiv
  e-prints

\bibitem[{{Torrey} {et~al.}(2012){Torrey}, {Vogelsberger}, {Sijacki},
  {Springel}, \& {Hernquist}}]{2012MNRAS.427.2224T}
{Torrey}, P., {Vogelsberger}, M., {Sijacki}, D., {Springel}, V., \&
  {Hernquist}, L. 2012, \mnras, 427, 2224

\bibitem[{{van de Weygaert}(1994)}]{1994A&A...283..361V}
{van de Weygaert}, R. 1994, \aap, 283, 361

\bibitem[{{Verhamme} {et~al.}(2006){Verhamme}, {Schaerer}, \&
  {Maselli}}]{2006A&A...460..397V}
{Verhamme}, A., {Schaerer}, D., \& {Maselli}, A. 2006, \aap, 460, 397

\bibitem[{{Vogelsberger} {et~al.}(2012){Vogelsberger}, {Sijacki}, {Kere{\v s}},
  {Springel}, \& {Hernquist}}]{2012MNRAS.425.3024V}
{Vogelsberger}, M., {Sijacki}, D., {Kere{\v s}}, D., {Springel}, V., \&
  {Hernquist}, L. 2012, \mnras, 425, 3024

\bibitem[{Voronoi(1908)}]{crll.1908.134.198}
Voronoi, G. 1908, Journal f{\"u}r die reine und angewandte Mathematik, 134, 198

\bibitem[{{Wang} {et~al.}(2013){Wang}, {Kaplan}, {Slane}, {Morrell}, \&
  {Kaspi}}]{2013ApJ...769..122W}
{Wang}, Z., {Kaplan}, D.~L., {Slane}, P., {Morrell}, N., \& {Kaspi}, V.~M.
  2013, \apj, 769, 122

\bibitem[{{Witt} \& {Gordon}(1996)}]{1996ApJ...463..681W}
{Witt}, A.~N. \& {Gordon}, K.~D. 1996, \apj, 463, 681

\bibitem[{{Witt} \& {Gordon}(2000)}]{2000ApJ...528..799W}
{Witt}, A.~N. \& {Gordon}, K.~D. 2000, \apj, 528, 799

\bibitem[{{Witt} {et~al.}(1992){Witt}, {Thronson}, \&
  {Capuano}}]{1992ApJ...393..611W}
{Witt}, A.~N., {Thronson}, Jr., H.~A., \& {Capuano}, Jr., J.~M. 1992, \apj,
  393, 611

\bibitem[{{Wolf}(2003)}]{2003CoPhC.150...99W}
{Wolf}, S. 2003, Computer Physics Communications, 150, 99

\bibitem[{{Wolf} {et~al.}(1998){Wolf}, {Fischer}, \&
  {Pfau}}]{1998A&A...340..103W}
{Wolf}, S., {Fischer}, O., \& {Pfau}, W. 1998, \aap, 340, 103

\bibitem[{{Wood} {et~al.}(2004){Wood}, {Mathis}, \&
  {Ercolano}}]{2004MNRAS.348.1337W}
{Wood}, K., {Mathis}, J.~S., \& {Ercolano}, B. 2004, \mnras, 348, 1337

\end{thebibliography}


\end{document}